\documentclass[10pt]{article}
\usepackage[centertags]{amsmath}
\usepackage{graphicx}
\usepackage{subfigure}
\usepackage[normal]{caption2}
\usepackage{epsfig}
\usepackage{amsfonts}
\usepackage{amssymb}
\usepackage{amsthm}
\usepackage{multirow}
\usepackage{newlfont}
\newlength{\defbaselineskip}
\newcommand{\setlinespacing}[1]%
           {\setlength{\baselineskip}{#1 \defbaselineskip}}

\newcommand{\Numbers}{\mathbb N}
\newcommand{\Field}{\mathbb F}

\newcommand{\ket}[1]{|{#1}\rangle}
\newcommand{\bra}[1]{\langle{#1}|}
\newcommand{\halfbra}[1]{\langle{#1}}

\newcommand{\sref}[1]{(\ref{#1})}
\newcommand{\Tr}{\mathrm{Tr}\,}

\newcommand{\abs}[1]{\left\vert#1\right\vert}

\theoremstyle{plain}
\newtheorem{thm}{Theorem}[section]
\newtheorem{cor}[thm]{Corollary}
\newtheorem{lem}[thm]{Lemma}
\newtheorem{prop}[thm]{Proposition}

\theoremstyle{definition}
\newtheorem{defn}{Definition}[section]
\theoremstyle{plain}
\newtheorem{rem}{Remark}[section]

\theoremstyle{assumption}
\newtheorem{assump}{Assumption}[section]

\makeatletter
\DeclareRobustCommand{\Cpp}
{\valign{\vfil\hbox{##}\vfil\cr
   \textsf{C\kern-.1em}\cr
   $\hbox{\fontsize{\ssf@size}{0}\textbf{+\kern-0.05em+}}\hbox{ }$\cr}%
}
\makeatother

\begin{document}

\title{Unconditional Security of Practical Quantum Key Distribution}
\author{H.J. Hupkes \\ Mathematical Institute of the University of Leiden}
\date{February 23rd 2004}

\maketitle

\begin{abstract}
We present an extension of the first proof for the unconditional security of the BB84
quantum key distribution protocol which was given by Mayers. We remove the constraint
that a perfect BB84 quantum source is required and the proof given here covers a
range of practical quantum sources. Nothing is assumed about the detector except that
the efficiency with which signals are detected is basis independent.
\end{abstract}



















\section{Introduction}
This paper presents an extension of the first proof for the unconditional security of a quantum key
distribution protocol, which was given by Mayers in \cite{MayersA}. The proof given here applies to a more general
class of quantum sources than the perfect single photon source analyzed in \cite{MayersA} and now covers a range of practical
quantum key distribution schemes.

The goal of any key distribution system is to allow two participants, typically called Alice and Bob, who initially
share no information, to share a secret random key at the end of the procedure. This secret key could then be used by both
Alice and Bob to encrypt messages they wish to send to each other through an insecure public channel they do not trust,
so that anybody who intercepts the encrypted message will learn nothing about the original message. There are many methods
available to encrypt messages, but they all require that Alice and Bob share a private key. As an example we mention the
classic Vernon one-time pad encryption scheme,
which requires Alice and Bob to share a private bit-string $k$ of length $n$ to encrypt a message $m$ containing $n$ bits.
Alice computes the encrypted message $m'$ via $m'[i] = m[i] \oplus k[i]$ and sends $m'$ to Bob, who finds $m$ by computing
$m'[i] \oplus k[i]$. An eavesdropper who intercepts the encrypted message $m'$ but has no information on the private key $k$ will learn
practically nothing about the message $m$.

If Alice and Bob agree to physically meet before exchanging any secret messages, it is of course very easy for them to generate
and share a secret bit-string. However, in the current information society in which there are millions of participants who wish
to communicate in a private manner, it is very impractical if not impossible for every pair of parties to meet and exchange keys.
One requires key distribution protocols in which all communication between Alice and Bob is public and can be monitored by
a potential eavesdropper Eve. However, after the protocol has terminated, Eve should know practically nothing about the key
which Alice and Bob share. At the moment, there are a number of classical key distribution systems which accomplish this task,
but they are only secure by virtue of the limited amount of computational power available to Eve. The classic RSA cryptosystem for
example relies upon the fact that it is extremely difficult to factorize products of two very large prime numbers.
The goal of a quantum key distribution system is to provide users the comforting idea that the security of the system
depends merely on the laws of nature and not on the unknown capabilities of adversaries.
With the possible rise of quantum computers which can factorize numbers in polynomial time, it can be argued that this
is not merely a theoretical issue.

A typical quantum key distribution protocol requires Alice to be in possession of a quantum source
and Bob to have a detection unit, which can perform some sort of measurement on the quantum states Alice sends.
In the BB84 protocol, which was proposed by Bennett and Brassard \cite{BB84MRef1}, Alice's source should be able
to produce photons linearly polarized at angles of exactly $0$, $\frac{\pi}{4}$, $\frac{\pi}{2}$ and $\frac{3\pi}{4}$.
Alice chooses secretly and randomly a string of basis-bits $a \in \{+ , \times \}^n$ and a string of key-bits $g \in \{0,1 \}^n = \Field_2^n$.
For every index $i$, Alice's source produces a photon polarized at $g[i] \frac{\pi}{2}$ if the corresponding basis bit was $+$
and a photon polarized at $\frac{\pi}{4} + g[i] \frac{\pi}{2}$ if the corresponding basis bit was $\times$.
Bob also chooses a secret string of basis bits $b \in \{ +, \times \}^n$ and measures the polarization of each photon sent by Alice
in the $+$ basis or the $\times$ basis, depending on $b$. In this way he determines a secret bit-string $h$
which reflects the outcome of his measurements. The key observation
is that if Bob and Alice share the same basis-bit for
some photon, then $g$ and $h$ will agree, while if their basis-bits differ, Bob will measure a zero or a one with equal probability.
By comparing their choice of basis $a$ and $b$ after the photon transmissions Alice and Bob can thus decide where $g$ and $h$ agree
and use this information to define a secret key.
Any potential eavesdropper Eve who intercepts the photons Alice sends to Bob has no information on the basis $a$ Alice is using
and thus cannot conclusively decide on Alice's key-bits $g$ by performing measurements on the photons. Even worse, if she
wishes to remain undetected, she must resend a photon to Bob, which in general will destroy the correlation between $g$ and $h$.
Intuitively, Alice and Bob can thus detect Eve with a large probability of success
by randomly choosing half of the exchanged photons where their bases $a$ and $b$
agree and revealing their key-bits $g$ and $h$ for these photons. If there are too many errors they should abort the protocol,
because Eve may know too much, while if the number of errors is small Eve knows nearly nothing and a key can safely be defined
using the remaining part of the photons.


The protocol we consider is a minor variant of the BB84 protocol. Since the introduction of this protocol in 1984,
a great deal of effort has been spent in order to prove that this protocol is secure against any attack
by Eve allowed by the laws of quantum physics. Many limited attacks were analyzed \cite{BB84MRef1, Mref2, Mref7, Mref8, Mref9, Mref10, Mref11, Mref12},
but it was only
in 1996 that Mayers provided the first proof of unconditional security \cite{MayersA}.
By now, Mayers argument has been followed up by other proofs of the security of ideal single-photon quantum key distribution \cite{ME7, ME8ShorPreskill}.
In particular, in \cite{ME8ShorPreskill} the authors relate the BB84 protocol to an entanglement purification protocol and give a conceptually
simple security proof.

We note here that unconditional security only means that there is no restriction on Eve's attack.
It thus does not mean that there is no condition on the apparatus used by Alice and Bob and
it is exactly this point that distinguishes the different security proofs now available.
The major advantage of the framework used in the Mayers proof is that it assumes nearly nothing about the
detector Bob uses, as opposed to e.g. the proof in \cite{ME8ShorPreskill}, which requires an ideal detector together with an ideal source.
Here the term ideal means that the equipment performs exactly as specified by the protocol.
In \cite{ImperfectDevices} a slight extension of the argument in \cite{ME8ShorPreskill} is used to analyze
slight deviations from the ideal source and detector, but there are still explicit assumptions on the source, channel and detector.
The weakness of the original Mayers proof is the assumption that the source emits perfectly aligned photons
at a rate of exactly one per pulse.
In practice, perfect single photon sources are not available and practical implementations use either dim laser pulses
or post-selected states from parametric downconversion. Unfortunately, both signal types contain multi-photon contributions
which might seriously compromise the security of quantum key distribution. In addition, there is always a slight spread in
the polarization axes of the emitted photons.

In \cite{MayersImp}, Mayers argument is extended to include multi-photon sources and it is shown that the security of BB84 is
maintained if the fraction of pulses that contain more than one photon is sufficiently small. This paper
deals with the issue of the imperfect polarization, by showing that the BB84 protocol remains secure if the deviation
from the perfect source is small, in a sense which we will make exact. We do not cover the multi-photon situation,
but we believe that it is merely a technicality to apply a similar extension of the type in \cite{MayersImp} to the proof
given here. Our proof follows closely the lines of \cite{MayersA, MayersImp} and makes use of the ideas contained therein.

This paper is organized as follows. In Section \ref{sc:protocol}, we define the variant of the BB84 protocol we will analyze. Section \ref{sc:qpsource}
introduces the notion of a quasiperfect source and discusses some practical types of quantum sources that are included by this
definition. We provide exact definitions for the concepts of privacy and security against tampering
in Section \ref{sc:mainresult} and use these definitions to state our main theorems. The technical proofs of these theorems
will be given in Section \ref{sc:proofmainresult}.

\section{The Protocol}
\label{sc:protocol}
In this section we define the variant of Bennett and Brassard's BB84 protocol we shall analyze.
Alice and Bob first together specify a number of parameters, then the quantum transmissions
take place and finally a classical negotiation is performed to define the key.

We employ a so-called randomizing box in the protocol, which is assumed to act independently of Alice and Bob
and whose functioning is trusted by both Alice and Bob. In particular, we shall assume that Eve cannot get
at the information in the box before it is announced and Eve cannot intercept the announcement
of the basis-bit to Bob in step \sref{l:qtBB}. The presence of this box is merely a technical convenience in the proof
and poses no real restriction on the protocol, since the box may simply be taken to be Bob's computer.
If Alice does not trust Bob's computer, she should not be exchanging a secret key with him in the first place.

The protocol requires that Alice is in possession of a quantum source which, given a basis-bit $a \in \{0, 1\}$ and a key-bit $g \in \{0, 1\}$,
produces some quantum state $\rho_a^g$, which need not necessarily be pure. Alice should also be able to send this
quantum state to
Bob along some quantum channel which is vulnerable to attack. In Section \ref{sc:mainresult} we shall introduce the constraints on the quantum source
and pre-agreement parameters which are necessary in order for the protocol to be private. However,
we shall assume nothing about the quantum channel or the measurement performed by Bob, except that Bob's detector
efficiency is basis-independent. Of course,
if the key distribution system is to be practical in a sense that Alice and Bob often share a key at the end of the protocol,
both Bob's equipment and the quantum channel will have to be adequate. The beauty of Mayers argument \cite{MayersA} is that these two issues
of privacy and usefulness are cleanly
separated from each other.

\subsubsection*{Pre-agreement}

  Alice and Bob together specify the following operating parameters.
  \begin{enumerate}
  \renewcommand{\labelenumi}{P\theenumi.}
  \makeatletter
  \renewcommand{\p@enumi}{P}
  \makeatother
    \item The length $m$ of the private key to be generated.
    \item The threshold $\delta_{\cal{P}} > 0$ for the error rate of the validation test.
    \item The number of bits $n > m$ which should be used for the validation test and for the key definition.
    \item A positive constant $\epsilon_N$ such that $N_{\mathrm{total}} = \lceil(4+ \epsilon_N)n \rceil$ is the number of quantum signals to be exchanged,
          where $\lceil x \rceil$ denotes the smallest integer which is at least as large as $x$.
    \item A security parameter $\epsilon > 0$, which directly determines the asymptotic security level of the protocol.
    \item A $r \times n$ binary parity check matrix $F$ for some integer $1 \le r \le n$ and a $m \times n$
              binary privacy amplification matrix $K$. See Appendix \ref{sc:aperr} for more information on parity check matrices.
  \end{enumerate}

\subsubsection*{Quantum Transmission}

  Alice and Bob repeatedly perform the following procedure, until the number of successfully exchanged photons is $N_{\mathrm{total}}$.
  \begin{enumerate}
  \renewcommand{\labelenumi}{QT\theenumi.}
  \makeatletter
  \renewcommand{\p@enumi}{QT}
  \makeatother
  \item Alice chooses randomly a basis-bit $a$ and a key-bit $g$.
  \item Alice announces to Bob that she is about to send a signal.
  \item Alice prepares the state $\rho_a^g$ and sends it to Bob.
  \item Alice announces the signal has been sent.
  \item Bob requests a basis-bit $b$ from the box if the previous one has been used.
        If Bob receives a signal, he performs a measurement on the received state, giving
        a value $h$. He informs Alice that the photon has been received and the number of successfully exchanged photons is incremented by one.
        If Bob does not receive a signal, he announces this and does nothing, retaining the basis-bit for future use. \label{l:qtBB}
  \end{enumerate}


\subsubsection*{Classical Negotiation}

  Alice and Bob go through the following steps and checks. If one of the checks is not passed, the protocol is aborted
  and Alice chooses her key $\vec{\kappa}$ randomly from $\Field_2^m$ in such a way that
  each $\vec{\kappa} \in \Field_2^m$ has equal probability to be chosen.
  \begin{enumerate}
  \renewcommand{\labelenumi}{C\theenumi.}
  \makeatletter
  \renewcommand{\p@enumi}{C}
  \makeatother
    \item The randomizing box announces Bob's basis $\vec{b} \in \Field_2^{N_{\mathrm{total}}}$ and a random set $R$ containing $N_{\mathrm{total}} / 2$ positions,
          which will be used for the verification test. \label{l:cAnnR}
    \item Bob announces $\vec{h}[R]$.
    \item The random box announces a permutation $\pi$ of the $N_{\mathrm{total}}$ elements.
    \item Alice announces her basis $\vec{a} \in \Field_2^{N_{\mathrm{total}}}$ and Alice and Bob calculate the set $\Omega = \{i \mid a[i] = b[i] \}$
          on which their bases agree.
    \item Alice and Bob check that the number of positions in $R$ on which Alice and Bob's basis agree is at least $n$
          and that the same holds for $\overline{R} = \{i \mid i \notin R \}$. \label{l:cCheckAgree}
    \item Let $S_{\cal{P}}$ be the set that contains the first $n$ positions in $\Omega \cap R$, where first refers to the ordering
          which results after applying the permutation $\pi$.
          Alice announces $g[S_{\cal{P}}]$. \label{l:cAnnG}
    \item Alice and Bob check that the number of differences $d_{S_\cal{P}}$ between $\vec{h}$ and $\vec{g}$ on
          $S_{\cal{P}}$ satisfies $d_{S_\cal{P}} \le \lfloor \delta_{\cal{P}} n \rfloor$,
          where $\lfloor x \rfloor$ denotes the largest integer $y$ satisfying $y \le x$. \label{l:cVerP}
    \item Let $S_{\cal{K}}$ be the set that contains the first $n$ positions in $\Omega \cap \overline{R}$, where
          first again refers to the ordering
          which results after applying the permutation $\pi$.
          Alice announces the syndrome $\vec{s} = F\vec{g}[S_{\cal{K}}]$ and defines the key $\vec{\kappa} = K\vec{g}[S_{\cal{K}}]$.
    \item Bob applies error correction to his bits $\vec{h}[S_{\cal{K}}]$ using the syndrome $\vec{s}$ to get $\vec{h}'[S_{\cal{K}}]$ and defines the
          key $\vec{\kappa}_B = K\vec{h}'[S_{\cal{K}}]$. See Appendix \ref{sc:aperr} for details on error correction.

  \end{enumerate}

\section{The source}
\label{sc:qpsource}


In the BB84 protocol one requires a source with takes as input a basis-bit $a$ and a key-bit $g$ and
produces a state $\rho_a^g$ over some finite dimensional Hilbert space $\cal{H}_Q$.
We shall prove the security of BB84 for a special class of sources
which we call quasiperfect sources. In this section we introduce and discuss this notion,
which is defined formally below.

\begin{defn}
A source which emits quantum states $\{\rho_a^g\}_{a=0, 1}^{g = 0, 1}$ over some finite Hilbert space $\cal{H}_Q$
is called quasiperfect with parameters $(\beta_{qp}, \gamma_{qp})$ if there exist
projection matrices $P_a^g$ and $\widetilde{P}_a^g$ for $a =0,1$ and $g=0,1$,
such that the following conditions hold.


\begin{enumerate}
\renewcommand{\labelenumi}{S\theenumi.}
\makeatletter
\renewcommand{\p@enumi}{S}
\makeatother
\item $ P_a^0 + P_a^{1} = \widetilde{P}_a^0 + \widetilde{P}_a^{1} = \mathbf{1}_{\cal{H}_Q}$ for $a=0,1$. \label{l:sResId}
\item
  We have the identity $\rho_0^0 + \rho_0^1 = \rho_1^0 + \rho_1^1$ and correspondingly define $H = \rho_0^0 + \rho_0^1$. \label{l:sDefH}
\item
  $\Tr P_a^g H = 1$ for $a =0, 1$ and $g = 0,1$. \label{l:sTrPH}
\item
  There exist unitary $T_a$ such that $T^{\dagger}_a P_a^g T = \widetilde{P}_a^g$ for $a = 0, 1$ and $g = 0,1$.
  In addition, $T^{\dagger}_a H T_a = H$ for $a =0,1$. \label{l:sDefT}
\item
  $\widetilde{P}_a^0 H \widetilde{P}_a^1 = 0$, for $a=0, 1$. \label{l:sOrthPT}
\item
  $\widetilde{P}_a^g \rho_{\overline{a}}^0 \widetilde{P}_a^g = \widetilde{P}_a^g \rho_{\overline{a}}^1 \widetilde{P}_a^g$,
  for $a=0,1$ and $g=0,1$. \label{l:sSplit}
\item There exist unitary $S_a$ such that
  $S_a^{\dagger} P_a^g S_a$ and $S_a^{\dagger} \rho_a^g S_a$ are diagonal for $a=0, 1$ and $g=0,1$. In particular, this means that
  $P_a^g$ and $\rho_a^g$ commute. \label{l:sDefS}
\item $\Tr P_a^g \rho_a^{\overline{g}} \le \beta_{qp}$ for $a=0,1$ and $g=0,1$.
\item Letting $\Lambda_a^g$ be the set of eigenvalues of the Hermitian matrix $P_a^g H - \widetilde{P}_a^g H$ and defining $\Delta_a^g = \sum_{\lambda \in \Lambda_a^g} \abs{\lambda}$,
      we have $\Delta_a^g \le \gamma_{qp}$ for $a=0,1$ and $g=0,1$. \label{l:sAbsTr}
\end{enumerate}  \qed
\end{defn}

The following lemma states some elementary properties of a quasiperfect source which follow directly from
the definitions given above.

\begin{lem}
\label{lm:TrivProp}
Consider a quasiperfect source with the corresponding matrices $\widetilde{P}_a^g$, $P_a^g$ and $H$ and let
$\Delta_a^g$ be defined as in \sref{l:sAbsTr}.
Then the following identities hold for all $a =0,1$ and $g=0,1$.
\begin{equation}
  \begin{array}{l}
  \widetilde{P}_a^0 \rho_{\overline{a}}^g \widetilde{P}_a^1 = - \widetilde{P}_a^0 \rho_{\overline{a}}^{\overline{g}} \widetilde{P}_a^1,\\
  \Tr P_a^0 \rho_a^1 = \Tr P_a^1 \rho_a^0, \\
  \Delta_a^0 = \Delta_a^1.
   \end{array}
\end{equation}
\end{lem}
\begin{proof}
The first identity follows immediately from properties \sref{l:sDefH} and \sref{l:sOrthPT}.
The last two identities follow immediately from  \sref{l:sDefH} and \sref{l:sResId}.
\end{proof}

In order to give some insight on the practical value of the above rather technical definition
of a quasiperfect source, we give two examples of such a source.
In particular, we show that our definition encompasses the ideal single-photon source analyzed in the Mayers proof
\cite{MayersA} and we give a nontrivial example which is very important for practical key distribution schemes.


We recall that an ideal BB84 source emits the states
$\rho_a^g = \ket{\Psi(a, g)}\bra{\Psi(a,g)}$ with
\begin{equation}
  \Psi(0, 0) = \left(\begin{array}{c}1 \\ 0 \end{array} \right), \, \,  \Psi(0, 1) = \left(\begin{array}{c}0 \\ 1 \end{array} \right)
\end{equation}
and
\begin{equation}
  \Psi(1, 0) = \frac{1}{\sqrt{2}}\left(\begin{array}{c}1 \\ 1 \end{array} \right), \, \,
  \Psi(1, 1) = \frac{1}{\sqrt{2}}\left(\begin{array}{c}1 \\ -1 \end{array} \right).
\end{equation}
Defining $I^0 = \left( \begin{array}{cc}1 & 0 \\ 0 & 0 \end{array} \right)$ and $I^1 = \mathbf{1}_2 - I^0$, we see that
\begin{equation}
\rho_0^g = I^g, \, \, \rho_1^g = R(\frac{\pi}{4})^{\dagger} I^g R(\frac{\pi}{4}),
\end{equation}
for $g=0,1$, where $R(\alpha)$ is the unitary rotation matrix with angle $\alpha$.

It is easy to see that this ideal source is also quasiperfect with  parameters $(0, 0)$,
by taking $P_a^g = \widetilde{P}_a^g = \rho_a^g$,
$S_0 = \mathbf{1}$ and $S_1 = R(-\frac{\pi}{4})$. 

We now give the nontrivial example of a quasiperfect source which can be seen as a generalization of the ideal
BB84 source. To do this, we consider probability distributions on the interval $[0, 2\pi]$.
If $p$ is such a distribution, we define the quantities
\begin{equation}
\begin{array}{lll}
s_p = \int_{0}^{2\pi} p(\alpha)\sin 2\alpha d\alpha,& \, \, \, &  c_p = \int_{0}^{2\pi} p(\alpha)\cos 2\alpha d\alpha, \\
s^{(2)}_p = \int_{0}^{2\pi} p(\alpha)\sin^2(\alpha) d\alpha, & \, \, \, & c^{(2)}_p = \int_0^{2\pi} p(\alpha) \cos^2(\alpha) d\alpha.
\end{array}
\end{equation}
For any angle $\phi$, we define the shifted distribution $p^{\phi}$ by
$p^{\phi}(\alpha) = p\big( (\alpha + \phi) \, \mathrm{mod} \, 2\pi\big)$.

\begin{thm}
\label{thm:QuasiPerfect}
Consider two probability distributions $p_0(\alpha)$ and $p_1(\alpha)$ on $[0, 2\pi]$ and
define the angles $\phi_a = \frac{1}{2} \arctan \frac{s_{p_a}}{c_{p_a}}$ for $a=0,1$.
Then the source which produces the states
\begin{equation}
\rho_a^g =  \int_{0}^{2\pi} p_a(\alpha) R(\alpha)^{\dagger}I^gR(\alpha) d\alpha
\end{equation}
is a quasiperfect source with parameters $(\beta_{qp}, \gamma_{qp})$, where
\begin{equation}
\begin{array}{lcl}
\beta_{qp} & = & \max \big( s^{(2)}_{p^{\phi_0}_0} , s^{(2)}_{p^{\phi_1}_1} \big), \\
\gamma_{qp} & = & \min \big( 2 \abs{\sin(\phi_1 - \phi_0 - \frac{\pi}{4})} , 2 \abs{\sin(\phi_0 - \phi_1 - \frac{\pi}{4})} \big).
\end{array}
\end{equation}
\end{thm}
\begin{proof}


We start by calculating
\begin{equation}
  R(\alpha)^{\dagger}I^gR(\alpha) =
  \left( \begin{array}{cc} \delta_{g0}\cos^2\alpha + \delta_{g1}\sin^2\alpha &
  (-1)^g \frac{1}{2} \sin2\alpha \\ (-1)^g\frac{1}{2}\sin2\alpha &
  \delta_{g0}\sin^2\alpha + \delta_{g1} \cos^2\alpha \end{array}\right).
\end{equation}
Now recalling that $\sin2(\alpha - \phi) = \sin2\alpha\cos(-2\phi) + \cos2\alpha\sin(-2\phi)$,
we see that
\begin{equation}
s_{p^{\phi_a}_a} = \cos(-2\phi_a) s_{p_a} + \sin(-2\phi_a) c_{p_a} = 0,
\end{equation}
by definition of $\phi_a$. This allows us to write
\begin{equation}
   \begin{array}{l}
   \rho_a^g = \int_{0}^{2\pi} p^{\phi_a}(\alpha )R(\alpha + \phi_a)^{\dagger}I^gR(\alpha+ \phi_a)d\alpha = \\
   = R(\phi_a)^{\dagger} \Big(\int_{0}^{2\pi} p^{\phi_a}(\alpha )R(\alpha)^{\dagger}I^gR(\alpha) d\alpha \Big) R(\phi_a)= \\
   R(\phi_a)^{\dagger} \left( \begin{array}{cc} \delta_{g0}c^{(2)}_{p^{\phi_a}_a} + \delta_{g1}s^{(2)}_{p^{\phi_a}_a} &
   0 \\ 0 & \delta_{g0}s^{(2)}_{p^{\phi_a}_a} + \delta_{g1} c^{(2)}_{p^{\phi_a}_a} \end{array}\right) R(\phi_a).
   \end{array}
\end{equation}

Now notice $\rho_0^0 + \rho_0^1 = \rho_1^0 + \rho_1^1 = \mathbf{1}_2 \equiv H$.
Defining $P_a^g = R(\phi_a)^{\dagger}I^gR(\phi_a)$ and $S_a = R(-\phi_a)$,
we immediately see that $S_a$ simultaneously diagonalizes $P_a^g$ and $\rho_a^g$ for $a=0,1$ and $g=0,1$.
It is also easy to see that $\Tr P_a^g \rho_a^{\overline{g}} = s^{(2)}_{p^{\phi_a}_a}$, which establishes
the claim about the parameter $\beta_{qp}$.
We also define
\begin{equation}
\begin{array}{l}
\label{eq:defnPTilde}
\widetilde{P}_0^g = R(\phi_1 -\frac{\pi}{4})^{\dagger}I^g R(\phi_1 - \frac{\pi}{4}), \\
\widetilde{P}_1^g = R(\frac{\pi}{4} + \phi_0)^{\dagger}I^g R(\frac{\pi}{4} + \phi_0). \\
\end{array}
\end{equation}
Since
\begin{equation}
  \left( \begin{array}{cc}1 & (-1)^g \\ (-1)^g & 1  \end{array} \right) \left(\begin{array}{cc}1 & 0
     \\ 0 & -1\end{array}\right)
  \left( \begin{array}{cc}1 & (-1)^g \\ (-1)^g & 1  \end{array} \right) = 0,
\end{equation}
one immediately verifies \sref{l:sSplit} by rotating the axis system over $-\phi_a$.
Condition \sref{l:sDefT} is satisfied if one defines $T_0 = R(\phi_1 - \frac{\pi}{4} - \phi_0 )$ and
$T_1 = R(\frac{\pi}{4} +\phi_0 - \phi_1 )$.
Finally, 
we calculate
\begin{equation}
 \Delta_a = R(-\phi_a)^{\dagger}(P_a^0 - \widetilde{P}_a^0) R(-\phi_a)  = \left(\begin{array}{cc}\sin^2(\psi_a) &
 -\frac{1}{2} \sin 2\psi_a \\ -\frac{1}{2} \sin 2\psi_a & - \sin^2(\psi_a) \end{array}\right),
\end{equation}
in which $\psi_0 = \phi_1 - \phi_0 - \frac{\pi}{4}$ and $\psi_1 = -\psi_0$.
Since the eigenvalues of $\Delta_a$ are $\pm \sin( \psi_a)$, the statement in the claim about the parameter $\gamma_{qp}$ immediately follows
using Lemma \ref{lm:TrivProp}, if we notice that in the definition \sref{eq:defnPTilde} we could have flipped the sign in front
of the angle $\frac{\pi}{4}$.


\end{proof}

\begin{rem}
If the probability distribution $p_a$ is symmetric around some angle $\alpha_a$, then $\phi_a = \alpha_a$.
The theorem shows how the parameters $(\beta_{qp}, \gamma_{qp})$ quantify the deviation of a quasiperfect source from the ideal BB84 source.
\end{rem}

The theorem above illustrates how a security proof which holds when a quasiperfect source
with small parameters $(\beta_{qp} , \gamma_{qp})$ is used will significantly
generalize the applicability of the original Mayers proof and will cover a range of practical quantum key distribution schemes.
In particular, since it is never possible in real life to perfectly align the polarization of the emitted photons,
the possibility to allow a small angular spread in these polarization axes is an essential element
of a practical security proof. We remark here that in Theorem \ref{thm:QuasiPerfect} we required
that the shape of the probability distribution which governs the alignment of the photon only depends on the basis-bit
and not on the key-bit. However, a convenient way to construct a source that satisfies property \sref{l:sDefH}
is to introduce an auxiliary system $A'$ with associated Hilbert space $\cal{H}_{A'}$. One then produces an entangled state $\rho_{AA'}$
and performs a measurement $M_a$, which depends only on the basis-bit $a$, acts only on the system $A'$ and has two possible outcomes.
If the key-bit $g$ is determined by the outcome of the measurement $M_a$, the shape of the probability distribution only
depends on the measurement $M_a$, which justifies the practicality of our assumption.

In this framework it is also possible to analyze the situation in which Eve performs a limited basis dependent attack,
as discussed in \cite{ImperfectDevices}.
This situation arises for example when we assume that Eve has supplied to Alice the source used for the quantum transmissions.
She could then have programmed the source to rotate the emitted photon slightly (relative to the ideal source)
if the corresponding basis bit was a $0$. She might even let the source vary the cheating strategy.
However, as long as Eve does not know during the quantum transmission phase which cheating tactic the
source is going to apply, it is sufficient to analyze the situation in which the source always emits the averaged state $\rho_a^g$.
We note here that this assumption means that Eve and the source do not share any non-constant
correlated random variables. This includes among others the absolute time and the number of already emitted photons.


Of course, the issue remains how one can test whether or not a source is quasiperfect and estimate the parameters.
In \cite{MayersImpApar}, the authors describe the issue of testing uncharacterized quantum equipment.
They show how to construct a so-called self-checking source which is guaranteed
to be a perfect BB84 source. However, their arguments assume that some specific probability distribution is known
exactly, which is of course never the case. We remark that it may be possible to adapt their argument
to include quasiperfect sources,
but we do not discuss this issue here.

\section{Main results}
\label{sc:mainresult}
In this section we state our main results, which concern the privacy and reliability of the BB84 protocol we discussed
in Section \ref{sc:protocol}. We shall consider the BB84 protocol in which a quasiperfect source with parameters $(\beta_{qp}, \gamma_{qp})$
is used and where in addition the conditions below hold.
\begin{assump}
\label{as:BB84Cond}
Let $\lambda$ be such that
\begin{equation}
  \label{eq:DefLambda}
  \frac{\lambda}{1-\lambda}\delta_{\cal{P}} \ge \frac{1}{2} \epsilon + \beta_{qp}.
\end{equation}
The minimal weight $d_w$ of linear combinations of rows from $F$ and $K$ which contain at least one row from
$K$ satisfies $d_w \ge 2(\frac{1}{1-\lambda}\delta_{\cal{P}} + \frac{1}{2} \gamma_{qp} + \epsilon)n$,
where the weight of a bit-string $\vec{v} \in \Field_2^N$ is defined to be the quantity $d(\vec{v},\vec{0})$, i.e. the number of
ones in $v$.
In addition, the matrix $F$ is the parity check matrix of a linear code which can correct $\lceil(\delta_{\cal{P}} + \epsilon)n\rceil$ errors.
Finally, Bob's detector efficiency is basis independent, i.e. the probability that a photon is successfully exchanged between Alice and Bob
is independent of the basis-bit used by Bob.
\end{assump}

Consider any possible attack by an eavesdropper Eve on the BB84 protocol. In general,
Eve will record all the classical messages announced by Alice and Bob
and perform a number of operations and measurements on the quantum states transmitted through the quantum
channel, possibly combined with measurements on auxiliary systems. Such an auxiliary system
could for example be a random number generator in order to introduce a certain randomness in the applied
eavesdropping tactic. After completion of all her operations, Eve will have acquired a vector $v$
of information of some kind, which we will consider to be an element in the set $\cal{V}$ of
all possible outcomes of her experiments.
We will consider the situation in which Eve has a fixed strategy for eavesdropping, that is,
if all the measurements on the external systems yield the same outcome and all the classical announcements
by Alice and Bob are the same, then Eve will perform the same operations and measurements on the emitted
quantum states. In this framework, the eavesdropping tactic employed by Eve defines a probability distribution
$P$ on the product space $\Field_2^m \times \cal{V}$, where $P(\vec{\kappa}, v)$ denotes the probability that
the key defined by Alice is $\vec{\kappa}$ and the information obtained by Eve is $v$.
If Alice and Bob want the key they share at the end of the protocol to remain secret,
then for any tactic employed by Eve the outcome $v$ should yield very little information about the key $\vec{\kappa}$.
This measure of correlation is conveniently expressed by the Shannon entropy $H_P(\vec{\kappa} \mid v)$, which is defined
as
\begin{equation}
H_P(\vec{\kappa} \mid v) = - \sum_{\vec{\kappa} \in \Field_2^m}\sum_{v \in \cal{V}} P(\vec{\kappa}, v)
\log_2 P(\vec{\kappa} \mid v).
\end{equation}
Here $P(\vec{\kappa} \mid v) = P(\vec{\kappa}, v) / P(v)$ denotes the conditional probability distribution of $\vec{\kappa}$ given $v$.
Note that in the ideal case the random variables $\vec{\kappa}$ and $v$ are independent, which means
$P(\vec{\kappa}, v) = P(\vec{\kappa})P(v) = 2^{-m} P(v)$, since each key $\vec{\kappa}$ is equally probable.
This immediately implies $H(\vec{\kappa} \mid v) = m$.

Our main result is expressed in the following theorem, which states that, under suitable operating conditions,
the maximal deviation from the ideal value of the conditional Shannon entropy that Eve can achieve
decreases exponentially as $n$ increases,
even if the rate of key generation
$m / n$ is kept at a constant level.

\begin{thm}
\label{thm:MainResult}
Consider the BB84 protocol in which a quasiperfect source with parameters $(\beta, \gamma)$ is used and
suppose that the conditions in Assumption \ref{as:BB84Cond} hold.
Consider any eavesdropping strategy that Eve can employ and let $\cal{V}$ be the set of all possible outcomes
of her measurements.
Denote by $P$ be the associated probability distribution
on the space $\Field_2^m \times \cal{V}$ for the random variable which gives jointly the key $\vec{\kappa} \in \Field_2^n$
defined by Alice and the information $v$ obtained by Eve. Then
there exist two functions $\epsilon_1(n, m, \epsilon)$ and $N(\epsilon)$, which are both independent of the strategy employed by Eve,
such that
\begin{equation}
\label{eq:MRPropE1}
H_P(\vec{\kappa} \mid v) \ge m - \epsilon_1(n, m ,\epsilon),
\end{equation}
for all $n \ge N(\epsilon)$.
Moreover, for any $\lambda > 0$, there exist constants $\mu(\lambda, \epsilon) > 0$ and $C(\lambda, \epsilon) > 0$,
such that
\begin{equation}
\label{eq:MRPropE2}
0 < \epsilon_1(n, m, \epsilon) \le C(\lambda, \epsilon) e^{-\mu(\lambda, \epsilon) n }
\end{equation}
for all $m$ which satisfy $m \le \lambda n$.
\end{thm}

The proof of this theorem will be given in subsequent sections. For the moment, we remark
that Corollary \ref{cr:asympGV} implies that the number of rows $r$ of a parity check matrix $F$ which meets
the conditions in Assumption \ref{as:BB84Cond}
can be chosen to satisfy $r/n \sim H_2\big(2(\delta_{\cal{P}} + \epsilon)\big)$, where $A(n) \sim B(n)$ means
$\lim_{n \to \infty} \frac{A(n)}{B(n)} = 1$ and $H_2$ is the binary entropy function
$H_2(x) = -\big(x \log_2 x + (1-x) \log_2 (1-x) \big)$. In view of this value for $r$,
Lemma's \ref{lm:binsum} and \ref{lm:privAmp} imply that we can choose a privacy amplification matrix $K$ with
$m$ rows that satisfies Assumption \ref{as:BB84Cond}, where
\begin{equation}
  \label{eq:asympRate}
  m /n \sim 1 - H_2\big(2 ( \delta_{\cal{P}} + \epsilon) \big) - H_2\big( 2( \delta_{\cal{P}} + \beta_{qp} + \frac{1}{2} \gamma_{qp} + \frac{3}{2} \epsilon) \big).
\end{equation}
To obtain this expression we have substituted $\frac{1}{1-\lambda}\delta_{\cal{P}} \approx \delta_{\cal{P}} + \frac{1}{2} \epsilon + \beta_{qp}$.
We can thus use the BB84 protocol to generate keys at the asymptotic rate $m /n$ given by \sref{eq:asympRate},
where the privacy level of the protocol increases as $n$ increases.
In the case where $\beta_{qp} = \gamma_{qp} = 0$, this means we can choose $\delta_{\cal{P}} = 5\%$
and still generate key-bits at a rate of $m /n \approx 6.2\%$.

We remark here that the rate \sref{eq:asympRate} is only a worst-case bound and is far from optimal.
In particular, if one relaxes the requirement that the error correcting code can correct all errors with weight less than
$(\delta_{\cal{P}} + \epsilon)n$ to the requirement that this can be done with probability exponentially close to one,
it is possible to choose $r/n \sim H_2(\delta_{\cal{P}} + \epsilon)$. Furthermore, in Remark \ref{rm:Conj}
we conjecture that it is possible to improve the third term in \sref{eq:asympRate}, which would lead to the bound
$m /n \sim 1-H_2(\delta_{\cal{P}}) - H_2( \delta_{\cal{P}} + \beta_{qp} + \frac{1}{2}\gamma_{qp})$, where we have taken $\epsilon \approx 0$.

It still remains to address the issue of the reliability of the BB84 protocol. In the situation that all the verification
tests succeed and Bob and Alice have both defined a key $\vec{\kappa}_B$ and $\vec{\kappa}$ respectively, they will need some
assurance that they indeed share the same private key. This is guaranteed by the following theorem.
\begin{thm}
Consider the BB84 protocol in which a quasiperfect source with parameters $(\beta, \gamma)$ is used and
suppose that the conditions in Assumption \ref{as:BB84Cond} hold.
Then there exists a function $\epsilon_2(n, \epsilon)$, bounded by $\epsilon_2(n, \epsilon) \le C(\epsilon)e^{-D(\epsilon) n}$
for some $C(\epsilon) > 0$ and $D(\epsilon) > 0$, such that
for any tactic employed by Eve, $P(\kappa \neq \kappa_B \cap \cal{P}) \le \epsilon_2(n, \epsilon)$,
in which $P(\kappa \neq \kappa_B \cap \cal{P})$ denotes the probability
that the keys defined by Bob and Alice are not equal while all the verification tests have succeeded.
\end{thm}
\begin{proof}
We consider the case where $\vec{a}$, $\vec{b}$, $\vec{g}$, $\vec{h}$ and $S = S_{\cal{P}} \cup S_{\cal{K}}$ are fixed but where
$R$ may still vary, that is, we do not know the partition of $S$ into $S_{\cal{P}}$ and $S_{\cal{K}}$.
We write $P'$ for the conditional probability distribution induced by this situation.
Since $R$ is uniformly distributed and is only announced after Bob has made his measurement to determine $h$,
each partition of $S$ is equally probable.
Let $E = d_S(\vec{g}, \vec{h})$ denote the total number of errors on $S$.
The error correcting code employed in the protocol can correct $\lceil(\delta_{\cal{P}} + \epsilon)n\rceil$ errors, which means
the keys defined by Alice and Bob will only differ if $d_{S_{\cal{K}}}(\vec{g}, \vec{h}) > \lceil(\delta_{\cal{P}} + \epsilon)n\rceil$, while
the test $\cal{P}$ only succeeds if $d_{S_{\cal{P}}}(\vec{g}, \vec{h}) \le \lfloor \delta_{\cal{P}} n \rfloor$.
First suppose that $E > \delta_{\cal{P}} n + (\delta_{\cal{P}} + \epsilon)n$. Then
\begin{equation}
P'(\kappa \neq \kappa_B \cap \cal{P}) \le 2^{-E} \sum_{i = 0}^{\lfloor\delta_{\cal{P}} n\rfloor} {E \choose i} \le e^{-2\big(\frac{\epsilon}{4}\big)^2E}
\le e^{-\frac{\epsilon^2}{4} (\delta_{\cal{P}} + \frac{1}{2}\epsilon)n} ,
\end{equation}
where we have used Corollary \ref{cr:binomialTail} with $p = \frac{1}{2}$ and
$t = \frac{1}{2} - \frac{\lfloor\delta_{\cal{P}} n\rfloor}{E} \ge \frac{\epsilon}{4}$.
Now suppose that $E \le \delta_{\cal{P}} n + (\delta_{\cal{P}} + \epsilon)n$. Then
\begin{equation}
P'(\kappa \neq \kappa_B \cap \cal{P}) \le 2^{-E} \sum_{i = \lceil(\delta_{\cal{P}} + \epsilon)n + 1 \rceil}^{E} {E \choose i}
  \le e^{-2 \big( \frac{1}{2} (\delta_{\cal{P}} + \frac{3}{2} \epsilon)n   \big) ^2  / E}
\le e^{- \frac{1}{4}\big( \frac{ (\delta_{\cal{P}} + \frac{3}{2} \epsilon)^2}{\delta_{\cal{P}} + \frac{1}{2}\epsilon}   \big)n} ,
\end{equation}
where we have used Lemma \ref{lm:binomialTail} with $p = \frac{1}{2}$ and
$t = \frac{\lceil(\delta_{\cal{P}} + \epsilon)n + 1 \rceil}{E} - \frac{1}{2} \ge \frac{\frac{1}{2}( \delta_{\cal{P}} + \frac{3}{2} \epsilon) n }{E}$.
Summing over all conditional probabilities $P'$ completes the proof.
\end{proof}

\section{Proof of Main Result}
\label{sc:proofmainresult}
In this section, we set out to prove our main result Theorem \ref{thm:MainResult}.
To do this, we first introduce two new protocols which differ from BB84, but for which it is
easier to analyze the attack by Eve.

\subsection{Reduction}

  We shall refer to the first modified protocol as BB84M. It consists of the following modifications to the BB84 protocol defined in Section
  \ref{sc:protocol}.
  \begin{itemize}
    \item Before the quantum transmission, the box announces to Alice through a completely secure channel the positions $R$.
    \item In step \sref{l:qtBB}, the randomizing box announces the bit $\widetilde{b}$ to Bob, defined by $\widetilde{b} = \overline{b}$
          if the position under consideration is in $\overline{R}$ and $\widetilde{b} = b$ otherwise.
    \item In step \sref{l:cAnnR}, the randomizing box announces $b$ for all positions, as usual.
  \end{itemize}
  Note that Bob doesn't know a priori which positions are in $R$, so during the transmission phase he will not
  know which basis-bit will be
  announced by the box.
  The intuitive idea behind this modification is that in this situation,
  Bob has measured in the wrong basis for all the positions in $S_{\cal{K}}$ and thus has no information about Alice's key.
  This modified procedure hence does not define a key distribution system, but is used only in the proof.
  In this light, we do not need to worry about the practicality of any of these modifications (for example, the private announcement of $R$
  by the box
  to Alice). All that is required is that in principle it is possible.
  The usefulness of this modified protocol is established by the following result.

  \begin{prop}
  \label{prp:red1}
  For any strategy adopted by a potential eavesdropper Eve, the random variable giving jointly Alice's private key and the information
  gathered by Eve has the same probability distribution in both protocols.
  \end{prop}
  \begin{proof}
    The only thing that has been changed is the announcement of the basis-bit from the randomizing box to Bob, but this
    cannot be intercepted by Eve due to the assumption on this box.
    Since Alice's choice for $\vec{a}$ and $\vec{g}$ are equivalent the emitted states are also equivalent. Since Bob's detector efficiency
    is basis independent, the subsets of photons which are successfully exchanged during the quantum transmission phase are equivalent.
    Also the information announced by Bob
    and the randomizing box is exactly the same, since on $R$ the outcome of Bob's measurement is unmodified.
    Alice does not use the announced string $R$ during the transition phase, so this makes no difference.
    So all the information which could be obtained by Eve, from either the quantum channel or the classical announcements,
    remains completely equivalent and thus the probability distributions are equal.
  \end{proof}

  We have seen that it is enough to prove the privacy in the modified protocol BB84M discussed above.
  However, if we can prove privacy in a
  further modified protocol BB84MM in which Eve receives more information than in the above protocol and
  can have a larger influence on the announcements,
  then this will immediately also imply privacy of the BB84M protocol and hence the original BB84 protocol.

  In particular, we shall consider BB84MM which consists of the following further modifications to BB84M.
  \begin{itemize}
    \item Alice generously announces $g[\overline{S}_{\cal{K}}]$ in step \sref{l:cAnnG} instead of merely $g[S_{\cal{P}}]$.
    \item Eve and Bob work together, that is, Bob tells Eve the announcement of the basis-bit he receives from the box and they together
          perform any measurement they want to determine a vector $\vec{h}$.
    \item Bob announces the complete vector $\vec{h}$ before the announcement of $R$ by the box in \sref{l:cAnnR}.
  \end{itemize}

The next proposition shows that it is indeed sufficient to prove the privacy of BB84MM against all possible attacks.
\begin{prop}
  \label{prp:red2}
  Consider any eavesdropping tactic Eve can employ on BB84M and let $P$ be the probability distribution of
  the resulting random variable which gives jointly Alice's key and the information gathered by Eve .
  Then there is a corresponding eavesdropping tactic on BB84MM with probability distribution $P'$ that
  satisfies $ H_{P'} \le H_{P}$.
\end{prop}
\begin{proof}
Notice that for any tactic on BB84M Eve can do exactly the same thing to eavesdrop on BB84MM, by letting
Bob perform the same measurement as in BB84M to get $h$. The only difference is that now Eve receives more
classical information than she did in BB84MM, i.e. $v' = (v, c_{extra})$, where $v'$ is Eve's information
in the BB84MM protocol and $v$ denotes the information gathered in the BB84M protocol.
We compute
\begin{equation}
\begin{array}{l}
H_{P'} = \sum_{\vec{\kappa} \in \Field_2^m}\sum_{v' \in \cal{V'}} P'(\vec{\kappa}, v') \log_2 \frac{1}{P'(\kappa \mid v')} = \sum_{\vec{\kappa} \in \Field_2^m}
\sum_{v \in \cal{V}} P(\vec{\kappa}, v)\sum_{c_{extra} \mid v}
  \frac{P'(\vec{\kappa}, v')}{P(\vec{\kappa}, v)} \log_2 \frac{P'(v')}{P'( \vec{\kappa}, v')} \le \\
\sum_{\vec{\kappa} \in \Field_2^m}\sum_{v \in \cal{V}} P(\vec{\kappa}, v) \log_2 \big( \sum_{c_{extra} \mid v} \frac{P'(\vec{\kappa}, v')}{P(\vec{\kappa}, v)}
\frac{P'(v')}{P'( \vec{\kappa}, v')}  \big) = H_{P},
\end{array}
\end{equation}
in which the inequality follows from Lemma \ref{lm:Jensen}. Here we have used the notation $\sum_{c_{extra} \mid v}$ to denote
the sum over all $c_{extra}$ for which $(v, c_{extra}) \in \cal{V}'$.

\end{proof}

Notice that this final reduction makes it possible to consider Eve and Bob as a single participant we shall call
Eve-Bob, who wishes to find out as much as possible about Alice's key.


\subsection{Formalism}
In this section we describe the formalism used to model Eve-Bob's attack on BB84MM.
The system seen by Eve-Bob can be seen as a state in a Hilbert space $\cal{H}_{sys} = \cal{H}_C \otimes
\cal{H}_S$, where $\cal{H}_C$ is a Hilbert space which describes
all the classical bit-strings generated during the protocol by Alice
and the randomizing box
and $\cal{H}_S = \bigotimes_{i=1}^{N_{\mathrm{total}}}{\cal{H}_Q}$ is the state space for the ensemble of transmitted quantum states.
We have $\cal{H}_C = \mathrm{span} \{ \ket{c}\}_{c \in C}$ for some set $C$ of states which we will define later.
Each state $c \in C$ will correspond to a classical bit-string and since these bit-strings can be perfectly distinguished
from one another, the corresponding states are all mutually orthogonal.

Any quantum state in a Hilbert space $\cal{H}$ is fully defined by the corresponding density matrix, which is
a Hermitian linear operator $\rho : \cal{H} \to \cal{H}$ that satisfies $\Tr \rho = 1$ and $(x, \rho x) \ge 0$
for all $x \in \cal{H}$. For finite dimensional Hilbert spaces such an operator is described by
a Hermitian non-negative matrix with unit trace.
The density matrix $\rho_{sys}$ for any state in $\cal{H}_{sys}$ encountered by Eve-Bob can be written
in the canonical form
\begin{equation}
  \rho_{sys} = \sum_{c \in V}P(c)\ket{c} \bra{c} \otimes \rho_c,
\end{equation}
where $V$ is a subset of $C$ and $P$ is a probability distribution on $V$, i.e. $P(c) \ge 0$ and $\sum_{c \in V}{P(c)} = 1$.

For notational convenience, we define the concept of a measurement operator, which will be used to describe measurements
on quantum systems.
\begin{defn}
A measurement operator on a Hilbert space $\cal{H}$ is a linear Hermitian operator $F: \cal{H} \to \cal{H}$ that satisfies
$(x,Fx) \ge 0$ for all $x \in \cal{H}$.     \qed
\end{defn}
The result of a general measurement on a system described in a Hilbert space $\cal{H}$ can be seen as an outcome of a random variable
$q$ reflecting the measured physical quantity. The probability distribution of the outcomes can be described using a positive
operator valued measure, defined below.
\begin{defn}
A positive operator valued measure (POVM) on a Hilbert space $\cal{H}$ consists of
a set of outcomes $Q$ together with a set  $\{ F_q \}_{q \in Q}$
of measurement operators on $\cal{H}$, such that
$\sum_{q \in Q}{F_q} = \mathbf 1_{\cal{H}}$.
For every outcome $q$ of the measurement, the probability of obtaining that outcome when performing the measurement on a
system with state $\rho$ is given by $\Tr F_q \rho$. \qed
\end{defn}
We note here that the POVM description can include measurements performed on external systems and possible
probes attached to the state $\rho$. We refer to \cite{POVM} for a general discussion on generalized measurements.

Eve-Bob's attack can be seen as a generalized measurement on the emitted state and thus can be described using the POVM formalism.
Actually, two measurements are performed: one before the classical announcements by Alice and the randomizing box and one after these announcements.
However, it is technically easier to describe the attack as a single POVM acting on the complete state $\rho_{sys}$. We will
use the restriction that the measurement of $\vec{h}$ is made before Alice and the box make their announcements to
derive a constraint on the form of the POVM.
To reflect the special nature of the classical announcements, we may assume that we can decompose every measurement operator
on $\cal{H}_{sys}$ as a sum of terms $\Pi^C \otimes E^S$, where $\Pi^C$ is a projection operator which can be written as
$\Pi^C = \sum_{c \in A}{\ket{c}\bra{c}}$ for some subset $A \subset C$ and $E^S$ is a measurement operator acting on the state
space $\cal{H}_S$ of the photons. Now we may always assume for Eve-Bob's POVM that each measurement operator consists of
a single term, as we can otherwise split the measurement operator in multiple operators. This gives more detailed information
than the original POVM and hence has a lower conditional Shannon entropy, as can be seen from the proof of Proposition \ref{prp:red2}. 

From now on, we will omit the vector sign on bit-strings if the distinction between a bit and a bit-string is clear from the context.
The set of all classical states is given by $C = \{ ( a, g, \pi, R, s ) \}$,
running over all possible combinations, noting that the syndrome $s$ is a function of all the other classical
variables. We consider the string $\widetilde{b}$ announced by the box to Eve-Bob to be fixed during our analysis,
which is why we do not include $b$ as part of the information in $C$ as it can be calculated given $R$.
As a further convenient restriction, we assume that the set $C$ contains only those classically generated bit-strings
that pass the verification test \sref{l:cCheckAgree}. Since the key chosen by Alice is perfectly uniformly distributed
if this test fails, it is possible to impose this restriction without loss of generality.

The classical announcements received by Eve-Bob are $y = ( a, g[\overline{S}_{\cal{K}}], R, s, \pi )$
and we define $\cal{Y}$ to be the set of all such announcements $y$ which are possible under the restriction that the
test \sref{l:cCheckAgree} passes.
The complete view $v$ that Eve-Bob gets from her measurements is given by $v = (y, \vec{h}, j)$, where $j$ describes
 any additional
information Eve-Bob can infer
out of her measurements.
Thus Eve-Bob's attack can be described by a POVM $\{F_v\}$ in which $F_v = \Pi^C_{y(v)} \otimes E^S_v$.
To reflect the fact that the measurement of $h$ is also a POVM and occurs without any knowledge of the classical outcomes, we may write
$\sum_{v \mid h}{F_v} = \mathrm{1}_{\cal{H}_C}\otimes E^S_h$ for some measurement operator $E^S_h$.
For convenience, we assume that the set $\cal{V}$ is finite, which is a reasonable assumption due to the nature
of any measuring device. However, it
is merely a technical issue to extend the argument given here to
infinite sets $\cal{V}$, so this discussion can be avoided.
Without loss of generality, we may also assume that $P(v) > 0$ for all $v \in \cal{V}$, since any view with $P(v) =0$ does not
contribute to the conditional Shannon entropy. Finally, we need only consider attacks for which $P(\cal{P}) > 0$,
where $\cal{P} \subset \cal{V}$ is the subset of $\cal{V}$ which consists of all views $v$ which pass the verification test \sref{l:cVerP}.
Indeed, if this condition is not satisfied, the protocol is trivially secure since
then the key that Alice chooses is independent of her interactions with Eve-Bob.
We summarize the above discussion by defining the concept of a normalized attack.
\begin{defn}
An attack by Eve-Bob on the BB84MM protocol is a normalized attack if it can be described by a POVM $\{F_v \}_{v \in \cal{V}}$
on $\cal{H}_{sys}$, for which the following identities hold.
\begin{enumerate}
\renewcommand{\labelenumi}{N\theenumi.}
  \makeatletter
  \renewcommand{\p@enumi}{N}
  \makeatother
\item Every $v \in \cal{V}$ can be written as $v = (y, h, j)$, for some $y \in \cal{Y}$ and $h \in \Field_2^{N_{\mathrm{total}}}$.
\item The set $\cal{V}$ is finite and $P(v) > 0$ for all $v \in \cal{V}$, where $P$ is the probability distribution induced by Eve-Bob's attack. In addition,
      $P(\cal{P}) > 0$.
\item For every $v \in \cal{V}$, the corresponding measurement operator can be decomposed as
  $F_v = \Pi^C_{y(v)} \otimes E^S_v$ for some measurement operator $E^S_v$ on $\cal{H}_S$.
\label{l:nKet}
\item For every $h \in \Field_2^{N_{\mathrm{total}}}$, we have $\sum_{v \mid h}{F_v} = \mathrm{1}_{\cal{H}_C}\otimes E^S_h$ for some measurement operator $E^S_h$ on $\cal{H}_S$. \label{l:nDefEH}
\end{enumerate} \qed
\end{defn}
The above discussion combined with Propositions \ref{prp:red1} and \ref{prp:red2} imply that once we have established the
following result, the proof of Theorem \ref{thm:MainResult} will be complete.
\begin{thm}
\label{thm:ToShow}
Consider the BB84MM protocol in which a quasiperfect source with parameters $(\beta_{qp}, \gamma_{qp})$ is used and
suppose that the conditions in Assumption \ref{as:BB84Cond} hold.
Then there exist a function $\epsilon_1(n, m, \epsilon)$ that satisfies equation \sref{eq:MRPropE2},
together with a function $N(\epsilon)$,
such that for any normalized attack on the BB84MM protocol \sref{eq:MRPropE1} holds for all $n \ge N(\epsilon)$.
\end{thm}

The next lemma states some very useful properties that the measurement operators satisfy and will be used
often throughout the proof of Theorem \ref{thm:ToShow}.

\begin{lem}
\label{lm:SimpIds}
Consider a normalized attack on the BB84MM protocol.
For every classical outcome $y \in \cal{Y}$, we have $\sum_{v \mid y}{E^S_v} = \mathbf{1}_{\cal{H}_S}$.
In addition, for every $y \in \cal{Y}$ and $h \in \Field_2^{N_{\mathrm{total}}}$, we have
$\sum_{v \mid (y,h) }{ E^S_v } = E^S_h$.
\end{lem}
\begin{proof}
The first identity can easily be seen by noting that for each $c \in C$, there is exactly one classical outcome $y$ which is compatible. We
refer to this outcome as $y(c)$.
Each projection matrix $\Pi^C_y$ is diagonal on the $\ket{c}$ basis, so we see that
$\bra{c}\Pi^C_{x}\ket{c} = \delta_{x,y(c)}$. But since $\sum_{v}{\bra{c}\Pi^C_{y(v)}\ket{c}E^S_v} = \mathbf{1}_{\cal{H}_S}$,
the identity immediately follows, using the fact that every $y$ has at least one compatible $c$.
The second identity can be proved similarly using \sref{l:nDefEH}.
\end{proof}


It will turn out to be very convenient to consider the view $z$ = $(h, a, R,g[R], \pi)$ which gives part of the information $v$
gathered by Eve-Bob. We write $z_c = (a, R, g[R], \pi)$ for the classical part of the view $z$,
together with $\cal{Z}$ and $\cal{Z}_c$ for the set of all possible views $z$ and $z_c$ respectively.
Upon calculating the measurement operator for this partial view, we find
\begin{equation}
  F_z = \sum_{v \mid z}{F_v} = \sum_{y \mid z_c} {\sum_{v \mid (y, h(z))}{F_v}} = \sum_{y \mid z_c}{ \sum_{ v \mid (y, h(z))}{\Pi^C_y \otimes E^S_v }}
    = \Big(\sum_{y \mid z_c} { \Pi^C_y} \Big) \otimes E^S_{h(z)} =  \Pi^C_{z_c} \otimes E^S_{h(z)},
\end{equation}
which expresses the nontrivial fact that the measurement operator for any $z$ remains a simple tensor product.
With similar reasoning as in the proof of Lemma \ref{lm:SimpIds} we may conclude that for any $z_c$,
\begin{equation}
\label{eq:SumZId}
\sum_{z \mid z_c} E^S_{h(z)} = \mathbf{1}_{\cal{H}_S}.
\end{equation}
%
%
%
The following lemma shows how we can reduce a trace over the complete space $\cal{H}_{sys}$ into a trace which runs merely over
the state space for the photons $\cal{H}_S$.

\begin{lem}
\label{lm:TraceReduction}
Consider a density matrix $\rho_{sys}$ of a state in $\cal{H}_{sys} = \cal{H}_C \otimes \cal{H}_{S}$ of the form
\begin{equation}
  \rho_{sys} = \sum_{c \in V} P(c) \ket{c}\bra{c} \otimes \rho_c,
\end{equation}
where $\rho_c$ is a density matrix of a state in $\cal{H}_S$ and $P$ is a probability distribution on $V$.
Consider a measurement operator of the form
\begin{equation}
 F = \Big( \sum_{c \in A} \ket{c} \bra{c} \Big) \otimes F^S,
\end{equation}
where $A \subseteq V$. Then for any linear operators $W^1$ and $W^2$ acting on $\cal{H}_S$, we have
\begin{equation}
  \Tr_{\cal{H}_{sys}} \big( F W^1 \rho_{sys} W^2 \big) = P(A) \Tr_{\cal{H}_S} (F^S W^1 \rho_{sys, A} W^2),
\end{equation}
where $\rho_{sys,A}$ is given by $\mathbf{1}_{\cal{H}_S} / \Tr \mathbf{1}_{\cal{H}_S}$ if $P(A) = 0$ and otherwise by
\begin{equation}
  \rho_{sys, A} = \frac{1}{P(A)} \sum_{c \in A}{ P(c) \rho_c}.
\end{equation}
\end{lem}
\begin{proof}
We have
\begin{equation}
  \Tr (F W^1 \rho_{sys} W^2 ) = \sum_{c \ in A}{\sum_{c' \in V} {P(c') \Tr ( \ket{c}\halfbra{c}\ket{c'}\bra{c'} )
  \Tr (F^S W^1 \rho_{c'} W^2 ) }}
\end{equation}
Noticing $\Tr ( \ket{c} \halfbra{c}\ket{c'}\bra{c'} ) = \delta_{c c'}$, we see that the above expression reduces to
\begin{equation}
\sum_{c \in A}{ P(c) \Tr (F^SW^1 \rho_c W^2) } = \Tr (F^S W^1 \sum_{c \in A}P(c) \rho_c W^2).
\end{equation}
From this the claim immediately follows.
\end{proof}


Let us consider the setting described in Theorem \ref{thm:ToShow}.
We define the function $g(n, \epsilon)= e^{- \epsilon^2 n} + e^{-\frac{1}{2}\epsilon^2n}$, which vanishes exponentially
as $n$ increases.
For any integer $N$ and any two bitstrings $b, w \in \Field_2^N$ we introduce the notation
$\widetilde{P}_{\vec{b}}^{\vec{w}} = \bigotimes_{i=1}^N P_{b[i]}^{w[i]}$.
We also define, for any $z \in \cal{Z}$ and any constant $\epsilon_{\cal{L}} > 0$, the projection operator
$\widetilde{\Pi}_0(z, \epsilon_{\cal{L}})$ via
\begin{equation}
  \widetilde{\Pi}_0(z, \epsilon_{\cal{L}}) = \sum_{\vec{w} \in W(z, \epsilon_{\cal{L}})} \widetilde{P}^{\vec{w}}_{\widetilde{\vec{b}}},
\end{equation}
where
$W(z, \epsilon_{\cal{L}}) = \{ \vec{w} \in \Field_2^{N_{\mathrm{total}}} \mid d_{S_{\cal{K}}}(\vec{w},
\vec{h}(z) ) \ge ( \frac{1}{1- \lambda}\delta_{\cal{P}} + \frac{1}{2} \gamma_{qp} + \epsilon_{\cal{L}} ) n  \}$,
in which $\lambda$ is defined by \sref{eq:DefLambda}.
Using the above definitions, we introduce the subset of views $\cal{L}_{\epsilon_{\cal{L}}} \subset \cal{P} \subset \cal{V}$, defined by
\begin{equation}
\label{eq:defSmallSphere}
\cal{L}_{\epsilon_{\cal{L}}} = \left\{ v \in \cal{P} \mid \Tr \left[ F_v \widetilde{\Pi}_0(z, \epsilon_{\cal{L}}) \rho
\widetilde{\Pi}_0(z, \epsilon_{\cal{L}}) \right] \le \sqrt{ g(n,\epsilon) } P(v) \right\}.
\end{equation}
In \cite{MayersA}, views $v \in \cal{L}$ were said to satisfy the small sphere property.
Our approach to proving Theorem \ref{thm:ToShow} will be to decompose the state emitted by the
source via $\rho = \big( (\widetilde{\Pi}_0 + (1 -\widetilde{\Pi}_0)\big) \rho \big(\widetilde{\Pi}_0 + ( 1- \widetilde{\Pi}_0)\big)$
and correspondingly split the expression $P(\kappa,v) = \Tr F_{\kappa,v} \rho$.
For views which satisfy the small sphere property we shall use the fact that $\Tr F_{\kappa,v} \widetilde{\Pi}_0 \rho \widetilde{\Pi}_0$
is small to bound the differences $P(\kappa_1,v) - P(\kappa_2,v)$, which proves that $v$ does not yield a significant amount of information
on the key $\kappa$. The following proposition roughly says that almost every view $v \in \cal{P}$ satisfies the small sphere property,
which makes it reasonable to assume that views which do not possess this property do not pose a large security threat.

\begin{prop}
\label{prp:SmallSphere}
Consider the BB84MM protocol in which a quasiperfect source with parameters $(\beta_{qp}, \gamma_{qp})$ is used and suppose
that the conditions in Assumption \ref{as:BB84Cond} hold.
Consider any normalized attack by Eve-Bob on BB84MM and let $P$ be the associated probability distribution.
%
Then $P(\cal{L_{\epsilon}}) \ge P(\cal{P}) - \sqrt{g(n, \epsilon)}$.
\end{prop}
\begin{proof}
We consider a slight variant of BB84MM, consisting of the following modifications.
\begin{itemize}
\item For each position in $\overline{R}$, Alice's source produces $\rho_{\bar{a}}$ instead of $\rho_{a}$.
\item For each position in $\overline{R}$, Alice applies the unitary transformation $T_{\bar{a}}^{\dagger} \rho T_{\bar{a}}$
to the photon, which makes it
diagonal in the $\widetilde{P}_{\bar{a}}$ measurement basis.
\item Alice performs a measurement on each photon before sending it to Eve-Bob.
For each position $i$ in $R$, Alice measures in the $P_{a[i]}$ basis, while for each position in $\overline{R}$,
Alice measures in the $\widetilde{P}_{\bar{a}[i]}$ basis. Alice records the results as $g_\cal{T}$ for future reference.
\end{itemize}
Let $\rho'$ denote the state emitted by the source in this modified protocol and
write $P'$ for the probability distribution defined by Eve-Bob's attack on this modified protocol.
For convenience, we define $\delta = \frac{\delta_{\cal{P}}}{1 - \lambda}$. Without loss of generality,
we shall assume that the first $\frac{N_{\mathrm{total}}}{2}$ positions belong to $R$ and the second $\frac{N_{\mathrm{total}}}{2}$ positions belong to $\overline{R}$.
We write $v_{\cal{T}} = (a, R, g_{\cal{T}})$ for the results received by Alice and let $\cal{V}_{\cal{T}}$ be
the set of all possible results Alice can receive.
We can then model the measurement of Alice as a POVM
\begin{equation}
  \left\{\Pi_{v_{\cal{T}}} = \Pi^C_{a,R} \otimes
  P(a, R, g_{\cal{T}}[R]) \widetilde{P}(\overline{a}, R, g_{\cal{T}}[\overline{R}]) \right\}_{ v_{\cal{T}} \in \cal{V}_{\cal{T}}},
\end{equation}
in which
\begin{equation}
\begin{array}{l}
P(a, R, g_{\cal{T}}[R]) = \bigotimes_{i=1}^{\frac{1}{2}N_{\mathrm{total}}} P_{a[i]}^{g_{\cal{T}}[i]} \otimes \mathbf{1}_{\cal{H}_S^{\otimes \frac{1}{2}N_{\mathrm{total}}}}, \\
\widetilde{P}(\overline{a}, R, g_{\cal{T}}[\overline{R}]) =  \mathbf{1}_{\cal{H}_S^{\otimes \frac{1}{2}N_{\mathrm{total}}}} \otimes \bigotimes_{i=\frac{1}{2}N_{\mathrm{total}} + 1}^{N_{\mathrm{total}}} \widetilde{P}_{\bar{a}[i]}^{g_{\cal{T}}[i]}
\end{array}
\end{equation}
Notice that in this case, each measurement operator is in fact a projection operator.
This allows us to compute the state seen by Eve-Bob after Alice's measurement, which is simply $\Pi \rho' \Pi$ if $\Pi$ is the
projection operator associated with the outcome received by Alice.

We now define a test $\cal{T}$, which is a function of Eve-Bob's announcement of $h$ and the results of Alice's
measurement $g_{\cal{T}}$. The test $\cal{T}$ succeeds if the number $d_{S_{\cal{P}}}(h, g_{\cal{T}})$ of differences between $h$ and $g_{\cal{T}}$
on $S_{\cal{P}}$ satisfies
$d_{S_{\cal{P}}}(h, g_{\cal{T}}) \le \delta n$, while the number of differences $d_{S_{\cal{K}}}(h, g_{\cal{T}})$ on $S_{\cal{K}}$
satisfies  $d_{S_{\cal{K}}}(h, g_{\cal{T}}) \ge (\delta + \frac{1}{2}\gamma_{qp} + \epsilon) n$.
Formally, we can consider $\cal{T}$ to be a subset of the combined view $\cal{Z} \times \cal{V}_{\cal{T}}$.
Letting $\cal{T}(z) \subseteq \cal{V}_{\cal{T}}$ be the set of Alice's views which pass the test given a value of $z$, we can write
$\cal{T} = \bigcup_{z \in \cal{Z}}{{z} \times \cal{T}(z)}$.
%
%
%
%
We can thus calculate
\begin{equation}
\label{eq:pcalT}
P'(\cal{T}) = \sum_{z \in \cal{Z}}{P'(z , \cal{T}(z))} = \sum_{z \in \cal{Z}}{P'(z \mid \cal{T}(z) ) P'(\cal{T}(z)) }.
\end{equation}
The interesting thing to note is that if $\cal{T}(z)$ is true, we know that the state after Alice's measurement is given by
$(\Tr \phi)^{-1} \phi$ where $\phi = \sum_{v_{\cal{T}} \in \cal{T}(z)}{ \Pi_{v_{\cal{T}}} \rho' \Pi_{v_{\cal{T}}} }$.
Now note that
\begin{equation}
  \sum_{v_{\cal{T}} \in \cal{T}(z)}{ \Pi_{v_{\cal{T}}} } = \Pi^C_{a, R} \otimes \Big( \sum_{g_{\cal{T}}
    \in \cal{G}(z) }
   {P(a,R,g_{\cal{T}}[R]) \widetilde{P}(\overline{a}, R,g_{\cal{T}}[\overline{R}])  } \Big) =
   \Pi^C_{a, R} \otimes \overline{\Pi}_1(z) \widetilde{\Pi}_0(z)  .
\end{equation}
in which $\cal{G}(z) = \{ g_{\cal{T}} \mid d_{S_{\cal{P}}}(g_{\cal{T}}, h) \le \delta n
   \wedge d_{S_{\cal{K}} }(g_{\cal{T}}, h) \ge (\delta + \frac{1}{2}\gamma_{qp} + \epsilon)n \}$ and
\begin{equation}
  \Pi_1(z) = \sum_{\vec{w} \in W_1(z)} P^{\vec{w}}_{\widetilde{\vec{b}}},
\end{equation}
where $W_1(z) = \{ w \in \Field_2^{N_{\mathrm{total}}} \mid d_{S_{\cal{P}}}(w, h(z) ) >  \delta n  \}$.
Here we have used that $\widetilde{b}[i] = \overline{a}[i]$ for every position $i \in S_{\cal{K}}$
and $\widetilde{b}[i] = a[i]$ for every position $i \in S_{\cal{P}}$, together
with the completeness condition \sref{l:sResId}.

The important observation now is that each emitted photon is diagonal on the basis it is measured in,
which allows us to write
\begin{equation}
\sum_{v_{\cal{T}} \in \cal{T}(z)}{ \Pi_{v_{\cal{T}}} \rho' \Pi_{v_{\cal{T}}} } = (\sum_{v_{\cal{T}} \in \cal{T}(z)}{ \Pi_{v_{\cal{T}}}}) \rho' (\sum_{v_{\cal{T}} \in \cal{T}(z)}{ \Pi_{v_{\cal{T}}}}).
\end{equation}
Now define the projection operators $\Pi_{\cal{T}(z)}$ and $\Pi^S_{\cal{T}(z)}$ via $\sum_{v_{\cal{T}} \in \cal{T}(z)}{ \Pi_{v_{\cal{T}}}} = \Pi_{\cal{T}(z)} = \Pi^C_{a, R, \pi} \otimes \Pi^S_{\cal{T}(z)}$
and note that
\begin{equation}
  \label{eq:defPiTZ}
  \Pi^S_{\cal{T}(z)} = \overline{\Pi}_1(z) \widetilde{\Pi}_0(z).
\end{equation}

Suppose now that Alice announces the result of her measurement $g_{\cal{T}}$ on each photon after Eve-Bob have announced $h$.
Given the partial outcome of Bob-Eve's measurement $h$, Eve-Bob can now announce
a string $R_{\mathrm{guess}}$, defined by $R_{\mathrm{guess}}[i] =1 \oplus g_{\cal{T}}[i] \oplus h[i]$,
where $\oplus$ denotes addition modulo two.
Denote by $\cal{T}' \subseteq \cal{Z} \times \cal{V}_{\cal{T}}$ the set of all events such that $R_{\mathrm{guess}}$ differs from $R$ on
$\cal{S}_{\cal{K}} \cup \cal{S}_{\cal{P}}$ on at most
$n(1 - \frac{1}{2}\gamma_{qp} - \epsilon)$ different positions, where we write $R[i] =1$ if $i \in R$ and $R[i] =0$ otherwise. Then it is the case that $\cal{T} \subseteq \cal{T}'$, since the total
number of differences between $R_{\mathrm{guess}}$ and $R$ for any combined view in $\cal{T}$ is bounded by
\begin{equation}
 d_{\cal{P}}(R_{\mathrm{guess}}, R) + d_{\cal{K}}(R_{\mathrm{guess}}, R) \le \delta n +  n(1 - (\delta + \frac{1}{2} \gamma_{qp} + \epsilon)) = n (1 - \frac{1}{2}\gamma_{qp} - \epsilon).
\end{equation}
Since Eve-Bob has no classical information when measuring $h$ and in particular does not know $R$, 
a correct announcement for $R_{\mathrm{guess}}[i]$ for some position $ i \in S_{\cal{P}} \cup S_{\cal{K}}$ corresponds
to a correct distinguishing of the state $P^{g_{\cal{T}}[i]}_{\widetilde{b}[i]} H$ from $\widetilde{P}^{g_{\cal{T}}[i]}_{\widetilde{b}[i]} H$.
Here we have used \sref{l:sTrPH}  and
$P^{g_{\cal{T}}[i]}_{\widetilde{b}[i]} H P^{g_{\cal{T}}[i]}_{\widetilde{b}[i]} = P^{g_{\cal{T}}[i]}_{\widetilde{b}[i]} H$,
together with a similar identity for $\widetilde{P}$, which both follow from \sref{l:sDefT}, \sref{l:sDefS} and the fact
that each emitted photon is diagonal on the basis it is measured in.

Theorem \ref{thm:Distinguish} in combination with \sref{l:sAbsTr} shows that that the success rate for a correct announcement
of $R_{\mathrm{guess}}$ on any $0 \le s \le 2n$ positions is bounded from above by $(\frac{1}{2} + \frac{1}{4} \gamma_{qp})^s$, since each position
in $S_{\cal{K}} \cup S_{\cal{P}}$ has probability $\frac{1}{2}$ to be in $R$.
Thus, even if Eve-Bob chooses the optimal strategy for determining $R_{\mathrm{guess}}$ which has
success rate $p_{\mathrm{suc}} = \frac{1}{2} + \frac{1}{4}\gamma_{qp}$, the probability $P'(\cal{T}')$
that $d_{S_{\cal{K}} \cup S_{\cal{P}}}(R_{\mathrm{guess}}, R) \le 2n  (\frac{1}{2}  - \frac{1}{2}\epsilon - \frac{1}{4} \beta)$
can be bounded from above by
\begin{equation}
P'(\cal{T}') \le \sum_{2n (\frac{1}{2} + \frac{1}{2} \epsilon + \frac{1}{4} \beta) \le i \le 2n} { 2n \choose i } p_{\mathrm{suc}}^i(1-p_{\mathrm{suc}})^{2n-i} \le e^{- \epsilon^2 n},
\end{equation}
in which we have used Lemma \ref{lm:binomialTail} with $t = \frac{1}{2} + \frac{1}{2}\epsilon + \frac{1}{4} \beta - p_{\mathrm{suc}} = \frac{1}{2}\epsilon$.
We thus obtain $P' ( \cal{T}) \le P' ( \cal{T}' ) \le e^{- \epsilon^2 n}$.

On the other hand, we can use \sref{eq:pcalT} to compute
\begin{equation}
  \begin{array}{lcl}
  P'( \cal{T}) & = &  \sum_{z \in \cal{Z}} { P\big( z \mid \cal{T}(z)\big) P\big(\cal{T}(z)\big) } \\
  & = &  \sum_{z \in \cal{Z}}{ \Big( \Tr F_z \Pi_{\cal{T}(z)} \rho' \Pi_{\cal{T}(z)} / ( \Tr \Pi_{\cal{T}(z)} \rho' ) \Big)
      \Big( \Tr \Pi_{\cal{T}(z)} \rho' \Big)} \\
  & = & \sum_{z \in \cal{Z}}{  \Tr F_z \Pi_{\cal{T}(z)} \rho' \Pi_{\cal{T}(z)} } = \sum_{z \in \cal{Z}}
  {  \Tr \Pi^C_{z_c} \otimes E^S_{h(z)} \Pi^S_{\cal{T}(z)} \rho' \Pi^S_{\cal{T}(z)} }. \\
  \end{array}
\end{equation}
We can now use Lemma \ref{lm:TraceReduction} to transform the above expression into
\begin{equation}
  P'(\cal{T}) = \sum_{z \in \cal{Z}} P'(z_c) \Tr (E^S_{h(z)} \Pi^S_{\cal{T}(z)} \rho'_{z_c} \Pi^S_{\cal{T}(z)} ),
\end{equation}

We remark here that the above expression resembles the definition of $\cal{L}$ in \sref{eq:defSmallSphere}, except
for the presence of the projection operator $\overline{\Pi}_1$. The idea is that for views in $\cal{P}$,
$\overline{\Pi}_1 \rho' \overline{\Pi}_1 \approx \rho'$ in some sense, since the number of errors on $S_{\cal{P}}$ is small.
We thus set out to bound the quantity
\begin{equation}
  \Delta = \sum_{z \in \cal{P}}{ \Tr \Big(F_z \Pi^S_{\cal{T}(z)} \rho' \Pi^S_{\cal{T}(z))}  - F_z \widetilde{\Pi}_0 \rho' \widetilde{\Pi}_0 \Big)}.
\end{equation}
We can use Lemma \ref{lm:TraceReduction} in combination with the expression \sref{eq:defPiTZ} for $\Pi^S_{\cal{T}(z)}$ to explicitly
split the sum over $z \in \cal{P}$ and write
\begin{equation}
  \begin{array}{l}
     \Delta = 
     \sum_{z_c \in \cal{Z}_c}\sum_{z \in V_{z_c} \cap \cal{P}}{ P(z_c) \Tr E^S_z \Big( \overline{\Pi}_1 \widetilde{\Pi}_0 \rho'_{z_c} \overline{\Pi}_1 \widetilde{\Pi}_0 - \widetilde{\Pi}_0 \rho'_{z_c} \widetilde{\Pi}_0  \Big)},
   \end{array}
\end{equation}
where $V_{z_c}=\{ z' \in \cal{Z} \mid z'_c = z_c \}$.

From now on, we shall consider $z_c$ to be fixed, so we consider each term in the sum above individually.
Note that $z_c$ contains information on the value $g$ for each bit in $R$.
We can thus write $\rho'_{z_c}$ as a tensor product (possibly after reordering bits)
$\rho'_{R} \otimes \rho'_{\overline{R}}$, in which $\rho'_{\overline{R}} =  2^{-\abs{R}} H^{\otimes \abs{R}}$.
Let $a = a(z_c)$ and define the unitary matrix $U_a$ which diagonalizes $\rho'_{z_c}$, $\widetilde{\Pi}_0$ and $\overline{\Pi}_1$ simultaneously.
Such a matrix exists due to our assumptions \sref{l:sDefS} and \sref{l:sDefT} on the source,
by letting $U_a = \bigotimes_{R} S_{a[k]} \otimes \bigotimes_{\overline{R}}
S_{\bar{a}[k]} T_{\bar{a}[k]}^{\dagger}$.
Note that $\overline{\Pi}_1$ operates only on $\rho'_{S_{\cal{P}}}$ while $\widetilde{\Pi}_0$ operates only on
$\rho'_{S_{\cal{K}}}$. In addition, since $U_a$ depends only on the classical part of $z$, we have for any
$z \in V_{z_c}$
\begin{equation}
  \begin{array}{l}
      \Tr E^S_z \Big( \overline{\Pi}_1 \widetilde{\Pi}_0 \rho'_{z_c} \overline{\Pi}_1 \widetilde{\Pi}_0 - \widetilde{\Pi}_0 \rho'_{z_c}
      \widetilde{\Pi}_0  \Big) = \\
      \Tr U_a^{\dagger}E^S_z U_a \Big( U_{a[R]}^{\dagger} (\mathbf{1} - \overline{\Pi}_1) U_{a[R]} U_{a[R]}^{\dagger} \rho'_{R} U_{a[R]} U_{a[R]}^{\dagger}
      (\mathbf{1}-\overline{\Pi}_1) U_{a[R]} \otimes \\
       U_{a[\overline{R}]}^{\dagger}   \widetilde{\Pi}_0 U_{a[\overline{R}]} U_{a[\overline{R}]}^{\dagger} \rho'_{\bar{R}} U_{a[\overline{R}]}
       U_{a[\overline{R}]}^{\dagger} \widetilde{\Pi}_0 U_{a[\overline{R}]}   \Big) = \\
      \Tr U_a^{\dagger}E^{S}_z U_a \Big( \Pi^d_1 \rho^{'d}_{R} \Pi^d_1 \otimes \widetilde{\Pi}^d_0 \rho^{'d}_{\overline{R}} \widetilde{\Pi}^d_0 \Big) 
   \end{array}
\end{equation}
where everything marked with a superscript $d$ has been diagonalized.

We now, independently of the non-classical part of $z$,  bound each diagonal element of $\Pi^d_1 \rho^{'d}_{R} \Pi^d_1 \otimes \widetilde{\Pi}^d_0 \rho^{'d}_{\bar{R}} \widetilde{\Pi}^d_0$.
%
%
%
Without loss (possibly rearrange matrix positions) we shall assume the identities
\begin{equation}
P_a^{d,a} = \left( \begin{array}{cc} \mathbf{1}_{d^a_a} & 0 \\ 0 & 0 \\ \end{array}\right), \, \, \, P_a^{d,\bar{a}} = \left( \begin{array}{cc} 0 & 0 \\ 0 & \mathbf{1}_{d^{\bar{a}}_a} \\ \end{array} \right),
\end{equation}
\begin{equation}
\rho_a^{d, a} = \mathrm{diag}( \alpha^a_{1, 2}, \ldots \alpha^a_{a, d_a^a}, \beta^a_{a, 1} , \ldots, \beta^a_{a, d_a^{\bar{a}}} ), \, \, \, \, \,
\rho_a^{d, \bar{a}} = \mathrm{diag}( \beta^{\bar{a}}_{a, 1}, \ldots \beta^{\bar{a}}_{a, d_a^{a}}, \alpha^{\bar{a}}_{a, 1} , \ldots \alpha^{\bar{a}}_{a, d_a^{\bar{a}}} ).
\end{equation}
For intuition purposes, we remark that the $\alpha$ values are in general large when compared to the $\beta$ values.
It is easy to see that (after rearranging), $\Pi^d_1 = \Pi^d_{1, S_{\cal{P}}} \otimes \mathbf{1}_{rest'}$.
Also write $\rho^{'d}_{R} = \rho^{'d}_{S_{\cal{P}}} \otimes \rho^{'d}_{rest'}$.
We now consider the diagonal $D^n \times D^{n}$  matrix $\Pi^d_{1, S_{\cal{P}}} \rho^{'d}_{S_{\cal{P}}} \Pi^d_{1, S_{\cal{P}}}$,
where $D$ is the dimension of the state space for a single photon $\cal{H}_Q$.
Let $w$ be a string in $\{0, 1, \ldots D -1 \}^{n}$ and let $e(w)$ be the corresponding $w$-th diagonal element of $\rho^{'d}_{S_{\cal{P}}}$, that is,
\begin{equation}
e(w) = \prod_{i = 1}^{n}{\big(\rho_{a_{S_{\cal{P}}}[i]}^{d,g_{S_{\cal{P}}}[i]}\big)_{w[i]w[i]} }.
\end{equation}
Similarly, we define $p(w)$ to be the $w$-th diagonal element of $\Pi_{1, S_{\cal{P}}}^d$.
Please note that $p(w) = 1$ if and only if $d_{S_{\cal{P}}}(w, h) > \delta n$.
Using the fact that the test $\cal{P}$ has passed, we know $d_{S_{\cal{P}}}(g, h) \le \delta_{\cal{P}} n = (1-\lambda) \delta n$.
Thus using the identity $d(w,g) \ge d(w, h) - d(h, g) \ge (\delta - (1- \lambda) \delta)n = \lambda \delta n$, we see that
if $p(w) = 1$ we must have $d_{S_{\cal{P}}}(w, g) \ge \lambda \delta n$.
Since this last inequality depends only on the value of $g$,
we obtain the following bound, which only depends on $z_c$,
\begin{equation}
p(w)e(w) p(w) \le \max(0, d_{S_{\cal{P}}}(w, g) - \lambda\delta n)  \prod_{i = 1}^{n}{\big(\rho_{a_{S_{\cal{P}}}[i]}^{d,g_{S_{\cal{P}}}[i]}\big)_{w[i]w[i]} } \equiv B(w).
\end{equation}
Now, 
noting that \sref{eq:SumZId} implies that for any $u \in \{0, \ldots, D-1\}^{N_{\mathrm{total}}}$
it holds that
$\sum_{z \in V_{z_c} \cap \cal{P}} (U_{a}^{\dagger}E^{S}_z U_a)_{uu} \le 1$,
one derives the inequality
\begin{equation}
\sum_{ z \in V_{z_c}}{T(z)} \le  \Tr \rho^{'d}_{rest'} \otimes \rho^{'d}_{\bar{R}} \sum_{w \in \{0, 1, \ldots D -1 \}^{ n}}{ B(w) } =  \sum_{w \in \{0, 1, \ldots D -1 \}^{ n}}{ B(w) }.
\end{equation}
Defining $V = \{ (a, b, c, d) \in \Numbers_0^4 \mid(a + b + c +d) \ge \lambda \delta n \}$ and
$W = \{0, 1, \ldots D-1\}^n$, we compute
\begin{equation}
\sum_{w \in W}{ B(w) } = \sum_{(i_0^0,i_0^1, i_1^0, i_1^1) \in V}
 \underset{\genfrac{}{}{0pt}{}{a = 0,1}{g =0,1}}{\prod}
{{N_a^g \choose i_a^g}
   \sum_{w \in \{1 , \ldots, d_a^{\bar{g}}\}^{i_a^g}}{\prod_{j=1}^{i_a^g}{\beta^g_{a, w[j]}} }
  \sum_{w' \in \{ 1, \ldots, d_a^g \}^{N_a^g - i_a^g}}{\prod_{j'=1}^{N_a^g - i_a^g}{\alpha^g_{a, w'[j']}} }},
\end{equation}
where $N_a^g$ is the number of positions in $S_{\cal{P}}$ that have basis-bit $a$ and key-bit $g$.
This can be seen by noting that given a choice of distance $d_{S_{\cal{P}}}(w,g) \ge \lambda \delta n$ and a distribution of the errors over the different bits,
which can occur with $\prod_{a, g}{{N_a^g \choose i_a^g}}$ possibilities, there are still $\prod_{a, g}{(d_a^{\bar{g}})^{i_a^g} (d_a^g)^{N_a^g - i_a^g}}$
compatible strings in $w$. Summing over the $e(w)$ values for these strings gives the above expression.
Now notice that for any $d, s \in \Numbers$ and any set of reals $\gamma_1, \ldots \gamma_d$, we have
\begin{equation}
  \sum_{w \in \{1, \ldots, d \}^s}{ \prod_{j=1}^s{ \gamma_{w[j]}} } = ( \sum_{j = 1}^{d}{ \gamma_j})^s.
\end{equation}
This can be easily seen by expanding the power. Using this, we obtain
\begin{equation}
  \begin{array}{l}
    \sum_{w \in  W}{ B(w) } = \sum_{(i_0^0,i_0^1, i_1^0, i_1^1) \in V}
    \underset{\genfrac{}{}{0pt}{}{a = 0,1}{g =0,1}}{\prod}
    {{N_a^g \choose i_a^g} ( \sum_{j = 1}^{d_a^{\bar{g}}}{\beta^g_{a, j}})^{i_a^g}
    ( \sum_{j = 1}^{d_a^{g}}{\alpha^g_{a, j}})^{N_a^g - i_a^g}} = \\
    \sum_{(i_0^0,i_0^1, i_1^0, i_1^1) \in V}
    \underset{\genfrac{}{}{0pt}{}{a = 0,1}{g =0,1}}{\prod}
    {{N_a^g \choose i_a^g} \beta_a^{i_a^g} (1- \beta_a)^{N_a^g - i_a^g}} = \\
    \sum_{(i_0, i_1) \in V'}{ {N_0 \choose i_0 }{ N_1 \choose i_1 } \beta_0^{i_0} (1 - \beta_0)^{N_0 - i_0} \beta_1^{i_1} ( 1- \beta_1) ^ {N_1 - i_1}  } \le e^{-\frac{1}{2} \epsilon^2 n},
  \end{array}
\end{equation}
where $V' = \{(i ,j) \in \Numbers_0^2 \mid i + j \ge \lambda \delta n \}$
and $\beta_a = \sum_{j = 1}^{d_a^{1}}{\beta^0_{a, j}} = \sum_{j = 1}^{d_a^{0}}{\beta^1_{a, j}}$.
This was obtained using Lemma \ref{lm:binomialTail}, with 
\begin{equation}
t = \lambda \delta - \max(\beta_0, \beta_1) = \frac{\lambda}{1-\lambda}\delta_{\cal{P}} - \max( \Tr P_0^1 \rho_0^1, \Tr P_1^0 \rho_1^0 ) \ge
\frac{\lambda}{1-\lambda}\delta_{\cal{P}} - \beta_{qp} \ge \frac{1}{2}\epsilon,
\end{equation}
where the last inequality follows from the assumption \sref{eq:DefLambda}.

Finally, this means we have obtained
\begin{equation}
  \Delta \le \sum_{z_c \in \cal{Z}_c }{P'(z_c) e^{-\epsilon^2n}} = e^{-\frac{1}{2} \epsilon^2 n}
\end{equation}
and hence
\begin{equation}
  \sum_{z \in \cal{P}} \Tr F_z \widetilde{\Pi}_0 \rho' \widetilde{\Pi}_0 = \sum_{z \in \cal{P}}
   P'(z_c) \Tr E^S_z \widetilde{\Pi}_0 \rho'_{z_c} \widetilde{\Pi}_0 \le e^{- \epsilon^2 n} + e^{-\frac{1}{2}\epsilon^2n}.
\end{equation}
We can now use the fact that since $z_c$ contains no information on $g[\overline{R}]$,
$\rho'_{z_c} = 2^{-\abs{\overline{R}}} \rho'_R \otimes H^{\otimes \abs{\overline{R}}}$.
However this also holds for the state emitted in the real protocol, so we have $\rho_{z_c} = \rho'_{z_c}$.
In addition, since the modifications do not influence the choice of $g[R]$, $a$, $R$ and $\pi$, we have $P'(g[R], a, R, \pi) = P(g[R],a, R, \pi)$.
This allows us to write
\begin{equation}
  \sum_{z \in \cal{P}} \Tr F_z \widetilde{\Pi}_0 \rho \widetilde{\Pi}_0 = \sum_{v \in \cal{V} \cap \cal{P}}  \Tr F_v  \widetilde{\Pi}_0 \rho\widetilde{\Pi}_0 \le g(n, \epsilon).
\end{equation}

We now employ Lemma \ref{lm:SmallProbability} with the probability distribution $P_{\cal{P}}(v) = P(v \mid \cal{P}) = P(v) / P(\cal{P}) $ on $\cal{P}$
and $q = P(\cal{P}) g(n, \epsilon)^{-\frac{1}{2}}$ to conclude that
\begin{equation}
P_{\cal{P}}(\cal{L}) \ge 1 - \sqrt{g(n, \epsilon))} / P(\cal{P}).
\end{equation}
The claim now follows upon multiplying both sides of the above identity by $P(\cal{P})$.
\end{proof}

\begin{cor}
  \label{cor:smallsphere}
 Suppose that $\epsilon_d \ge \epsilon$.
 Then
   $P(\cal{L}_{\epsilon_d}) \ge P(\cal{P}) - \sqrt{g(n, \epsilon)}$.
\end{cor}
\begin{proof}
The proof of Proposition \ref{prp:SmallSphere} goes through if we replace $\epsilon$ by $\epsilon_d$ everywhere.
Since $g(n, \epsilon_d) \le g(n, \epsilon)$, we see that
\begin{equation}
\left\{ v \in \cal{P} \mid \Tr \left[ F_v \widetilde{\Pi}_0(z, \epsilon_d) \rho
\widetilde{\Pi}_0(z, \epsilon_d) \right] \le \sqrt{ g(\epsilon_d, n) } P(v) \right\} \subseteq \cal{L}_{\epsilon_d}
\end{equation}
and hence
$P(\cal{L}_{\epsilon_d}) \ge 1 - \sqrt{g(n, \epsilon_d)} \ge 1- \sqrt{g(n, \epsilon)},$
which establishes the claim.
\end{proof}

The following proposition will be used to extract the key independent part of the probability distribution $P(\kappa,v)$.
The assumption \sref{l:sSplit} on the source plays a crucial role in the proof.
\begin{prop}
\label{prp:Ind}
Consider a BB84 source that is quasiperfect with parameters $(\beta_{qp}, \gamma_{qp})$.
Let $F$ be an arbitrary $r \times n$ binary matrix and $K$ be a $m \times n$ binary matrix,
for some integers $m$,$r$ and $n$ which satisfy $0 \le m, r \le r + m \le n$.
Define $d_w$ to be the minimal weight of linear combinations of rows from $F$ and $K$ which contain
at least one row from $K$.
Suppose that two arbitrary strings $b,h \in \Field_2^n$ and a constant $d''$ are given, such that $d'' \le \frac{1}{2}d_w$.
Let $X$ be a measurement operator acting on $\cal{H}^{\otimes n}_Q$ such that $X \widetilde{P}^{\vec{j}}_{\vec{b}} = 0$ for all
strings $j \in \Field_2^n$ which satisfy $d(h,j) \ge d''$.
For any $\kappa \in \Field_2^m$ and $s \in \Field_2^r$, define the set
\begin{equation}
  C_{\kappa, s} = \{g \in \Field_2^n \mid Fg = s \hbox{ and } Kg = \kappa\}
\end{equation}
and the state
\begin{equation}
\rho_{\kappa, s, \bar{b}} = \frac{1}{\abs{C_{\kappa, s}}}\sum_{g \in C_{\kappa, s} }  \bigotimes_{k = 1}^n{ \rho^{g[k]}_{\bar{b}[k]} }. 
\end{equation}
Then $\Tr X \rho_{\kappa, s, \bar{b}}$ is independent of $\kappa$.
\end{prop}
\begin{proof}
It is enough to show that for any two keys $\kappa, \kappa' \in \Field_2^m$ and
$\Delta \rho = \rho_{\kappa, s, \bar{b}} - \rho_{\kappa', s, \bar{b}}$, we have
\begin{equation}
  \label{eq:thmIndClaimToShow}
\widetilde{P}^{k}_b \Delta \rho \widetilde{P}^{l}_b = 0
\end{equation}
for all strings $k, l \in \Field_2^n$ which satisfy $d(k, l) < d_w$.
Indeed, assuming this, write
\begin{equation}
  \begin{array}{l}
    \Tr X \Delta \rho = \sum_k \sum_l \Tr X \widetilde{P}_b^k \Delta \rho \widetilde{P}_b^l = \sum_{k, l \mid d(k, l) \ge d_w}
      \Tr X \widetilde{P}_b^k \Delta \rho \widetilde{P}_b^l \\
    = \sum_{k, l \mid d(k, l) \ge d_w}  \Tr \widetilde{P}_b^l X \widetilde{P}_b^k \Delta \rho. \\
   \end{array}
\end{equation}
It can be seen that for every pair of strings $k,l \in \Field_2^n$ with $d(k,l) \ge d_w$, either $X \widetilde{P}_b^k= 0$ or $\widetilde{P}_b^l  X= 0$.
Indeed, assuming the contrary, then
$d(k, h) < \frac{1}{2}d_w$ and also $d(l, h) < \frac{1}{2}d_w$.
However $d(k,l) \le d(k,h) + d(h, l) < d_w$, which immediately gives a contradiction.
This fact now implies $\Tr X \Delta \rho=0$, which is the claim stated in the lemma.

We thus set out to show \sref{eq:thmIndClaimToShow}.
Using \sref{l:sSplit} and Lemma \ref{lm:TrivProp}, we can define the matrices
\begin{equation}
  \alpha_{b}^a = \widetilde{P}^a_{\bar{b}} \rho^0_b \widetilde{P}^a_{\bar{b}} = \widetilde{P}^a_{\bar{b}} \rho^1_b \widetilde{P}^a_{\bar{b}}
\end{equation}
and
\begin{equation}
  \beta_b^a=\widetilde{P}^a_{\bar{b}} \rho^0_b \widetilde{P}^{\bar{a}}_{\bar{b}} = - \widetilde{P}^a_{\bar{b}} \rho^1_b \widetilde{P}^{\bar{a}}_{\bar{b}} .
\end{equation}
With these definitions, for any four bits $b, d, e, f \in \Field_2$ we can define the matrix $V_b^{d, e, f}$ by
\begin{equation}
   V_b^{d, e, f} =  \widetilde{P}^d_{\bar{b}} \rho_b^e \widetilde{P}^f_{\bar{b}} =
(\beta_b^d)^{ d \oplus f } (-1)^{ (d \oplus f)e } (\alpha_b^d) ^ {d \oplus f \oplus 1},
\end{equation}
where $\oplus$ denotes addition modulo two.
We extend this definition to bit-strings $b,d,e,f \in \Field_2^n$ by writing
$\cal{V}_b^{d, e, f} = \bigotimes_{i=1}^{n} V_{b[i]}^{d[i], e[i], f[i]}$.
Since $V_b^{d, \bar{e}, f} = (-1)^{d \oplus f} V_b^{d, e, f}$,
we obtain the following identity for any bit-string $e' \in \Field_2^n$
\begin{equation}
\cal{V}_b^{d, e \oplus e', f} = (-1)^{e' \cdot (d \oplus f)}  \cal{V}_b^{d, e , f}.
\end{equation}

Let $G$ be the matrix
\begin{equation}
  G = \left( \begin{array}{c} K \\ F\end{array} \right)
\end{equation}
and write $x = (\kappa, s) \in \Field_2^{r + m}$ and $\rho_x = \rho_{\kappa, s, \bar{b}}$.
Let $C_x$ be the set of $g \in \Field_2^n$ which satisfies $Gg = x$.
Notice that indeed $C_x = C_{\kappa, s, \bar{b}}$ and that
for every $g \in C_x$, one can write $C_x = g \oplus C_0$.
Defining $(\rho_x)_{kl} = \widetilde{P}^k_{\bar{b}} \rho_x \widetilde{P}^l_{\bar{b}}$ and
fixing any $\theta \in C_x$, we calculate
\begin{equation}
  \label{eq:thmIndShiftX}
  \begin{array}{l}
    (\rho_x)_{kl} = \widetilde{P}^{k}_{\bar{b}}  \rho_x \widetilde{P}^{l}_{\bar{b}} = \frac{1}{\abs{C_x}} \sum_{g \in C_{x}} \cal{V}_b^{k, g, l} =
    \frac{1}{\abs{C_x}} \sum_{g \in C_{0}} \cal{V}_b^{k, g \oplus \theta, l} =                    \\
     (-1)^{\theta \cdot (k \oplus l)}  \frac{1}{\abs{C_x}} \sum_{g \in C_{0}} \cal{V}_b^{k, g \oplus \theta, l} =
     (-1)^{\theta \cdot (k \oplus l)} \widetilde{P}^{k}_{\bar{b}}  \rho_0 \widetilde{P}^{l}_{\bar{b}} =
     (-1)^{\theta \cdot (k \oplus l)} (\rho_0)_{kl}.
  \end{array}
\end{equation}

The above identity shows that it is sufficient to compute $(\rho_0)_{kl}$, which we therefore set out to do.
Write $\abs{C_0} = 2^q$, where $q$ is the dimension of $C_0$ and let $\{ \theta_1, \ldots , \theta_q \}$ be $q$ linearly independent
bit-strings which span $C_0$. For $ 0 \le j \le q$, let $C^{(j)}$ be the span of the strings $\theta_1, \ldots \theta_j$
and $\rho^{(j)} = \frac{1}{\abs{C^{(j)}}} \sum_{g \in C^{(j)}} \rho(g, \bar{b})$, in which
$\rho(g, \bar{b}) = \bigotimes_{i=1}^n \rho_{\bar{b}[i]}^{g[i]}$. Notice that $\rho^{(q)} = \rho_0$ and $\rho^{(0)} = \rho(0, \bar{b})$.
We shall prove by induction that for all $0 \le j \le q$, the following identity holds
\begin{equation}
(\rho^{(j)})_{kl} = \left\{ \begin{array}{ll} \cal{V}_b^{k, 0, l} & \hbox{if } (k \oplus l) \in C^{(j) \perp},
  \\ 0 & \hbox{otherwise}. \end{array}  \right.
\end{equation}
The $j=0$ case is trivial in view of the definition of $\cal{V}$ and the fact that $C^{(0)\perp} = \Field_2^n$.
Now, $C^{(j+1)} = C^{(j)} \cup (C^{(j)} \oplus \theta_{j+1})$, so
\begin{equation}
  (\rho^{(j+1)})_{kl} = \frac{1}{2}( \rho^{(j)})_{kl} ( 1 + (-1)^{ (k \oplus l) \cdot \theta_{j+1}} ).
\end{equation}
Note that $C^{(j+1)\perp} = C^{(j)\perp} \cap \{\theta_{j+1}\}^{\perp}$.
Observe also that if $(\rho^{(j+1)})_{kl} \neq 0$, we must have that $k \oplus l \in C^{(j)\perp}$ and
$(k \oplus l) \cdot \theta_{j+1} = 0\, \mathrm{mod} \, 2$,
which precisely means that $k \oplus l \in C^{(j+1)\perp}$. In this case, we see that $(\rho^{(j+1)})_{kl} = (\rho^{(j)})_{kl}$, which
concludes the induction argument.

Now, using \sref{eq:thmIndShiftX}, we see that for every $\theta \in C_x$, we have
$(\rho_x)_{kl} = (-1)^{(k \oplus l)\cdot \theta} (\rho_0)_{kl}$.
Also, every string $j \in C_0^{\perp}$ can be written as a unique linear combination of rows of $G$,
i.e. there exists a function $\lambda$ with $\lambda(j) \cdot G = j$ for every $j$ in $C_0^{\perp}$.
We can thus write, using $G\theta = x$,
\begin{equation}
  (\rho_x)_{kl} = (-1)^{\lambda(k \oplus l)\cdot x} (\rho_0)_{kl}.
\end{equation}
We are now ready to complete the proof.
We know that if $d(k, l) = w(k \oplus l) < d_w$ and $k \oplus l \in C_0^{\perp}$,
then we must have by definition of $d_w$ that $k \oplus l$ is a sum of rows of $F$ only.
This however means that $\lambda(k \oplus l) \cdot (\kappa, s)$ is independent of $\kappa$,
which immediately establishes the claim.
\end{proof}
\begin{rem}
\label{rm:Conj}
We conjecture that it is possible to generalize the argument above, if we assume that the probability
of a random linear combination of rows from $K$ and $F$ that contains at least one row from $K$
having weight smaller than $d_w$ is exponentially small.
We should then obtain $\Tr X \rho_{\kappa, s, \bar{b}} = t_{v} + \eta_{\kappa,v}$,
where $t_v$ is independent of $\kappa$ and $\eta_{\kappa,v}$ is exponentially small.
This result is enough to complete the privacy proof in a similar manner as described below.
\end{rem}

For any normalized attack by Eve-Bob on BB84MM, we
can calculate the probability distribution $P(\kappa,v)$ by considering the POVM which corresponds to the hypothetical
scenario in which Alice announces her key $\kappa$
after the protocol is completed.
Since the key $\kappa$ is revealed only after the complete protocol has finished and the measurement performed on the photons thus
cannot depend on it, this POVM can be seen to satisfy
$F_{\kappa,v} = \Pi^C_{y(v), \kappa} \otimes E^S_v$.
We thus calculate
\begin{equation}
P(\kappa,v) = \Tr F_{\kappa,v} \rho = P( \kappa,y) \Tr E^S_v \rho_{\kappa,y}.
\end{equation}
For ease of notation, we can reorder indices and write $\cal{H}_S = \cal{H}_S^{\cal{K}} \otimes \cal{H}_S^{\overline{\cal{K}}}$,
where $\cal{H}_S^{\cal{K}}$ is the state space of all the photons in the set $S_{\cal{K}}$ on which the key is defined.
We can also split $\rho_{\kappa,y} = \rho^{\cal{K}}_{\kappa,y} \otimes \rho^{\bar{\cal{K}}}_v$ correspondingly.
We can then use Lemma \ref{lm:PartialTrace} to define $E^{\cal{K}}_v = \Tr_{\cal{H}_S^{\bar{\cal{K}}}} E^S_v \rho_{ \kappa,y}$, which only
depends on $v$, and one can check
\begin{equation}
P(\kappa,v) = P(\kappa,y) \Tr E^{\cal{K}}_v \rho^{\cal{K}}_{\kappa,y},
\end{equation}
where the trace now runs over $\cal{H}_S^{\cal{K}}$.

For any nonnegative operator $X$ on $\cal{H}_S^{\cal{K}}$ and for any $y \in \cal{Y}$ and $\kappa \in \Field_2^m$, we define the ratio
\begin{equation}
  r_{\kappa,y}(X) = \frac{\Tr X \rho^{\cal{K}}_{\kappa,y} }{\Tr X \rho^{\cal{K}}_y },
\end{equation}
with the convention that $r_{\kappa,y}(X) = 1$ whenever the expression above is undefined.
It is easy to see that $r_{\kappa,y}(X) \ge 0$ and
\begin{equation}
  \label{eq:sumr}
  \sum_{\kappa \in \Field_2^m} r_{\kappa,y}(X) =2^m.
\end{equation}

The following proposition shows that for any view $v$ which satisfies the small sphere property, the
joint probabilities $P(\kappa,v)$ for all keys $\kappa$ are very similar and hence $v$ does not leak
a significant amount of information about the key.
\begin{prop}
\label{prp:ProbSplit}
Consider the BB84MM protocol in which a quasiperfect source with parameters $(\beta_{qp}, \gamma_{qp})$ is used and suppose
that the conditions in Assumption \ref{as:BB84Cond} hold.
Consider any normalized attack by Eve-Bob on BB84MM and let $P$ be the associated probability distribution.
Let $\epsilon_w$ be such that $\frac{1}{2}d_w = (\frac{1}{1-\lambda}\delta_{\cal{P}} + \frac{1}{2} \gamma_{qp} + \epsilon_w)n$
and note that $\epsilon_w \ge \epsilon$.
Consider any view $v$ in $\cal{L}_{\epsilon_w}$ and write $\widetilde{\Pi}_0 = \widetilde{\Pi}_0(z, \epsilon_w)$. Then
\begin{equation}
P(\kappa, v) = \pi_v + \eta_{\kappa,v},
\end{equation}
in which $\pi_v$ is a constant independent of $\kappa$ and $\eta_{ \kappa,v}$ is bounded according to
\begin{equation}
\eta_{ \kappa,v} \le 2^{-m}P(v) \Big(r_{ \kappa,y}(E^{\cal{K}}_v) +  r_{\kappa,v}(\widetilde{\Pi}_0 E^{\cal{K}}_v  \widetilde{\Pi}_0 ) \Big)
h(n, \epsilon) ,
\end{equation}
where $h(n, \epsilon) = 2g(n, \epsilon)^{\frac{1}{4}} + g(n, \epsilon)^{\frac{1}{2}}$.
\end{prop}
\begin{proof}
Note that due to the fact that the rows of $K$ and $F$ are linearly independent and each value of $g[S_{\cal{K}}]$ is equally probable,
$P(\kappa,y) = 2^{-m}P(y)$ and $\sum_{\kappa \in \Field_2^m}\rho^{\cal{K}}_{\kappa, y} = 2^{m} \rho^{\cal{K}}_y$.
This gives us
\begin{equation}
P( \kappa,v) = 2^{-m}  P(y) \Tr E^{\cal{K}}_v \rho^{\cal{K}}_{\kappa, y}.
\end{equation}
Write $\widetilde{\overline{\Pi}}_0$ for $1 - \widetilde{\Pi}_0$.
Using the identity
\begin{equation}
\begin{array}{l}
X = (A + \overline{A})X(A + \overline{A}) = \overline{A}X\overline{A} + AX(\overline{A} + \frac{1}{2}A)
+ (\overline{A} + \frac{1}{2}A)XA = \\ \overline{A}X\overline{A} + AX(I - \frac{1}{2}A) +(I - \frac{1}{2}A)XA =
\overline{A}X\overline{A} + AX + XA - AXA,
\end{array}
\end{equation}
we obtain
\begin{equation}
\begin{array}{lcl}
2^m \frac{1}{P(y)}P(\kappa,v) & = & \Tr E^{\cal{K}}_v \widetilde{\overline{\Pi}}_0 \rho^{\cal{K}}_{\kappa, y} \widetilde{\overline{\Pi}}_0  \\
& + & \Tr E^{\cal{K}}_v \widetilde{\Pi}_0 \rho^{\cal{K}}_{\kappa, y} + \Tr E^{\cal{K}}_v \rho^{\cal{K}}_{\kappa, y} \widetilde{\Pi}_0 \\
& - & \Tr E^{\cal{K}}_v \widetilde{\Pi}_0 \rho^{\cal{K}}_{\kappa, y} \widetilde{\Pi}_0.
\end{array}
\end{equation}
Proposition \ref{prp:Ind} implies that $\Tr E^{\cal{K}}_v \widetilde{\overline{\Pi}}_0 \rho^{\cal{K}}_{\kappa, y} \widetilde{\overline{\Pi}}_0$
is independent of $\kappa$, so we define
\begin{equation}
\begin{array}{l}
\pi_v = 2^{-m} P(y) \Tr E^{\cal{K}}_v \widetilde{\overline{\Pi}}_0 \rho^{\cal{K}}_{\kappa, y} \widetilde{\overline{\Pi}}_0, \\
\eta_{\kappa,v} = 2^{-m} P(y) \Big( \Tr E^{\cal{K}}_v \widetilde{\Pi}_0 \rho^{\cal{K}}_{\kappa, y} + \Tr E^{\cal{K}}_v \rho^{\cal{K}}_{\kappa, y} \widetilde{\Pi}_0
 -  \Tr E^{\cal{K}}_v \widetilde{\Pi}_0 \rho^{\cal{K}}_{\kappa, y} \widetilde{\Pi}_0 \Big). \\
\end{array}
\end{equation}

We now make the decomposition $E^{\cal{K}}_v = \sum_{l} \ket{ \phi^{\cal{K}}_{l,v}} \bra{\phi^{\cal{K}}_{l,v}}$.
Noting that the first two terms of $\eta_{ \kappa,v}$ are complex conjugates, we obtain
\begin{equation}
\begin{array}{l}
\abs{\eta_{ \kappa,v}}  \le   2^{-m} P(y) \Big( 2\sum_l\abs{ \bra{\phi^{\cal{K}}_{l,v}} \widetilde{\Pi}_0 \rho^{\cal{K}}_{\kappa, y} \ket{\phi^{\cal{K}}_{l,v}}  }
+ \Tr E_v^{\cal{K}} \widetilde{\Pi}_0 \rho_{\kappa, y} \widetilde{\Pi}_0 \Big).
\end{array}
\end{equation}
Since $\rho^{\cal{K}}_{\kappa, y}$ is a nonnegative hermitian matrix, we may employ the Cauchy-Schwartz inequality to write
\begin{equation} \begin{array}{l}
\abs{ \bra{\phi^{\cal{K}}_{l,v}} \widetilde{\Pi}_0 \rho^{\cal{K}}_{\kappa, y} \ket{\phi^{\cal{K}}_{l,v}}  } =
\abs{ \bra{\phi^{\cal{K}}_{l,v}} \widetilde{\Pi}_0 (\rho^{\cal{K}}_{\kappa, y})^{\frac{1}{2}} (\rho^{\cal{K}}_{\kappa, y})^{\frac{1}{2}} \ket{\phi^{\cal{K}}_{l,v}} } \le \\
\bra{\phi^{\cal{K}}_{l,v}} \widetilde{\Pi}_0 \rho^{\cal{K}}_{\kappa, y} \widetilde{\Pi}_0 \ket{\phi^{\cal{K}}_{l,v}}^\frac{1}{2}
  \bra{\phi^{\cal{K}}_{l,v}}  \rho^{\cal{K}}_{\kappa, y} \ket{\phi^{\cal{K}}_{l,v}}^\frac{1}{2}.
\end{array}
\end{equation}
Another application of Cauchy-Schwartz yields
\begin{equation}
\begin{array}{l}
\sum_l\abs{ \bra{\phi^{\cal{K}}_{l,v}} \widetilde{\Pi}_0 \rho^{\cal{K}}_{\kappa, y} \ket{\phi^{\cal{K}}_{l,v}}  }
\le \big(\sum_l \bra{\phi^{\cal{K}}_{l,v}} \widetilde{\Pi}_0 \rho^{\cal{K}}_{\kappa, y} \widetilde{\Pi}_0 \ket{\phi^{\cal{K}}_{l,v}}\big)^{\frac{1}{2}}
\big(\sum_l \bra{\phi^{\cal{K}}_{l,v}}  \rho^{\cal{K}}_{\kappa, y} \ket{\phi^{\cal{K}}_{l,v}}\big)^{\frac{1}{2}} = \\
\big(\Tr E_v^{\cal{K}} \widetilde{\Pi}_0 \rho_{\kappa, y}^{\cal{K}} \widetilde{\Pi}_0\big)^{\frac{1}{2}}
\big(\Tr E_v^{\cal{K}} \rho_{\kappa, y}^{\cal{K}}\big)^{\frac{1}{2}},
\end{array}
\end{equation}
and thus

\begin{equation} \begin{array}{l}
\abs{\eta_{\kappa,v}} \le 2^{-m}P(y) \big( \Tr E_v^{\cal{K}} \widetilde{\Pi}_0 \rho_{\kappa, y}^{\cal{K}} \widetilde{\Pi}_0 \big)^{\frac{1}{2}}
\big( (\Tr E_v^{\cal{K}} \widetilde{\Pi}_0 \rho_{\kappa, y}^{\cal{K}} \widetilde{\Pi}_0)^{\frac{1}{2}} +
  2(\Tr E_v^{\cal{K}} \rho_{\kappa, y}^{\cal{K}})^{\frac{1}{2}} \big) = \\
2^{-m}P(y) \Big( r_{\kappa, y}(\widetilde{\Pi}_0 E^{\cal{K}}_v \widetilde{\Pi}_0 ) \Tr E^{\cal{K}}_v \widetilde{\Pi}_0 \rho^{\cal{K}}_{y} \widetilde{\Pi}_0  \Big)^{\frac{1}{2}}
\Big( r_{\kappa, y}(\widetilde{\Pi}_0 E^{\cal{K}}_v \widetilde{\Pi}_0 )^{\frac{1}{2}} \Tr E^{\cal{K}}_v \widetilde{\Pi}_0 \rho^{\cal{K}}_{y} \widetilde{\Pi}_0
  ^{\frac{1}{2}} +
  2r_{\kappa, y}( E^{\cal{K}}_v)^{\frac{1}{2}}(\Tr E^{\cal{K}}_v  \rho^{\cal{K}}_{y}) ^{\frac{1}{2}} \Big) \le \\
  2^{-m} \max\{ r_{\kappa, y}(E^{\cal{K}}_v), r_{\kappa, y}(\widetilde{\Pi}_0 E^{\cal{K}}_v  \widetilde{\Pi}_0 ) \}
  \big( P(y)\Tr E^{\cal{K}}_v \widetilde{\Pi}_0 \rho^{\cal{K}}_{y } \widetilde{\Pi}_0
  \big)^{\frac{1}{2}}
\big( \big(P(y) \Tr E^{\cal{K}}_v \widetilde{\Pi}_0 \rho^{\cal{K}}_{y} \widetilde{\Pi}_0 \big)^{\frac{1}{2}} +
  2 \big(P(y) \Tr E^{\cal{K}}_v  \rho^{\cal{K}}_{y} \big)^{\frac{1}{2}} \big)
\end{array}
\end{equation}
We now use the identity $P(y) \Tr E^{\cal{K}}_v  \rho^{\cal{K}}_{y}  = \Tr F_v \rho  = P(v)$
together with the fact that $v \in \cal{L}_{\epsilon_w}$ to obtain the bound
\begin{equation}
\begin{array}{l}
\abs{\eta_{\kappa,v}} \le 2^{-m} \Big(r_{\kappa, y}(E^{\cal{K}}) +  r_{\kappa, y}(\widetilde{\Pi}_0 E^{\cal{K}}_v \widetilde{\Pi}_0 ) \Big)
  (\sqrt{g(n, \epsilon)} P(v))^{\frac{1}{2}} \Big(( \sqrt{g(n, \epsilon)} P(v))^{\frac{1}{2}} + 2 P(v)^{\frac{1}{2}} \Big) =   \\
  2^{-m} P(v) \Big(r_{\kappa, y}(E^{\cal{K}}) +  r_{\kappa, y}(\widetilde{\Pi}_0 E^{\cal{K}}_v \widetilde{\Pi}_0 ) \Big)
  \Big( 2g(n, \epsilon)^{\frac{1}{4}} +  g(n, \epsilon)^{\frac{1}{2}} \Big),
\end{array}
\end{equation}
which concludes the proof.

\end{proof}

\noindent We now have all the ingredients which are necessary to complete the privacy proof.

\begin{proof}[Proof of Theorem \ref{thm:ToShow}.]
Define $\epsilon_w$ and $\widetilde{\Pi}_0$ as in the statement of Proposition \ref{prp:ProbSplit}.
Fix a view $v \in \cal{L}_{\epsilon_w}$ and a real number $q \ge 1$.
For convenience, define
\begin{equation}
 a_{\kappa,v} = r_{\kappa,v}(\widetilde{\Pi}_0  E^{\cal{K}}_v \widetilde{\Pi}_0  ) + r_{\kappa,v}( E^{\cal{K}}_v ) .
\end{equation}
Note that
\begin{equation}
P(v) = \sum_{\kappa \in \Field_2^m} P(\kappa, v) = 2^m \pi_v + \sum_{\kappa \in \Field_2^m} \eta_{ \kappa,v},
\end{equation}
and thus recalling \sref{eq:sumr}
\begin{equation}
  \abs{P(v) - 2^m \pi_v } \le \sum_{\kappa \in \Field_2^m} \abs{ \eta_{ \kappa,v} } \le 2^{-m} P(v) h(n, \epsilon)
  \Big(\sum_{\kappa \in \Field_2^m} a_{\kappa,v}\Big)
  = 2P(v)h(n,\epsilon).
\end{equation}
From this we obtain the bound
\begin{equation}
\abs{P(\kappa \mid v) - \frac{1}{2^m}} = \frac{1}{P(v)}\abs{P(\kappa, v) - \frac{1}{2^m} P(v)} \le \frac{1}{P(v)}\Big(\abs{P(\kappa, v)
  - \pi_v} + \abs{\pi_v - \frac{1}{2^m} P(v)} \Big) \le \frac{1}{2^m}h(n, \epsilon)(a_{ \kappa,v} + 2).
\end{equation}
Recalling that $\sum_{\kappa \in \Field_2^m}a_{\kappa, v} = 2^{m+1}$, we see that
the set $\cal{K}_v = \{ \kappa \in \Field_2^m \mid a_{ \kappa,v} < 2q \}$ has at least $2^m(1 - \frac{1}{q})$ elements.
Thus defining the set $\cal{I}  = \Field_2^m \times \bigcup_{v \in \cal{L}_{\epsilon_w}} \{v\} \times \cal{K}_v \subseteq \cal{V}$,
we see that for all $( \kappa,v) \in \cal{I}$,
\begin{equation}
\label{eq:idOnI}
\abs{P(\kappa \mid v) - \frac{1}{2^m}} \le \frac{1}{2^m}(2q + 2)h(n, \epsilon).
\end{equation}
Now, since Alice chooses her key uniform randomly when the test $\cal{P}$ is not passed,
we have
\begin{equation}
H(\kappa \mid v) = -\sum_{\kappa, v} P(\kappa,v) \log_2 P(\kappa \mid v) \ge mP(\overline{\cal{P}}) - \sum_{( \kappa,v) \in \cal{I}} P(\kappa, v) \log_2 P(\kappa \mid v),
\end{equation}
where the inequality was obtained by noting that $\log_2 p \le 0$ for all $0 \le p \le 1$ and
that $\cal{I} \subseteq \cal{P} \times \Field_2^m$.
Writing $P(\kappa \mid v) = \frac{1}{2^m}(1 + \xi_{\kappa, v}) \ge 0$, where $\abs{\xi_{\kappa, v}} \le  (2q+2)h(n, \epsilon)$, noting
that $C (1+x) \log_2 (1 + x) \le C(1+x)\frac{\abs{x}}{\ln 2}$ for any $x \ge -1$ and using $P(\cal{I}) \le 1$, we see
\begin{equation}
H(\kappa \mid v) \ge mP(\overline{\cal{P}}) - \sum_{ ( \kappa,v) \in \cal{I}} P(\kappa, v) (-m + \frac{(2q+2)h(n,\epsilon)}{\ln 2}) =
m\big((P(\overline{\cal{P}}) + P(\cal{I}) \big) - \frac{(2q+2)h(n, \epsilon)}{\ln 2}.
\end{equation}
Using \sref{eq:idOnI} it is easy to see that
\begin{equation}
\begin{array}{l}
P(\cal{I}) = \sum_{v \in \cal{L}_{\epsilon_w}}P(v) \sum_{k \in \cal{K}_v}P(\kappa\mid v) \ge
P(\cal{L}_{\epsilon_w})(1 - \frac{1}{q})(1 - (2q+2)h(n, \epsilon)).
\end{array}
\end{equation}
From the above identity, we conclude
\begin{equation}
\begin{array}{l}
H(\kappa \mid v) \ge
m\Big(P(\overline{\cal{P}}) + (1-\frac{1}{q})(1 - (2q+2)h(n, \epsilon))(P(\cal{P})-P(\cal{P} \cap \overline{\cal{L}}_{\epsilon_w}))\Big) -\frac{(2q+2)h(n, \epsilon)}{\ln 2} \\
\ge m  - \frac{m}{q} -(m + \frac{1}{\ln2 } )(2q+2)h(n,\epsilon) -m P(\cal{P} \cap \overline{\cal{L}}_{\epsilon_w}).
\end{array}
\end{equation}
Now choose $q = \sqrt{\frac{m}{2(m+ \frac{1}{\ln2 }) h(n , \epsilon) } }$.
It is easy to see that there exists a function $N(\epsilon)$ which depends only on $\epsilon$, such that
$q \ge 1$ for all $n \ge N(\epsilon)$. Thus, for all $n \ge N(\epsilon)$, we have
\begin{equation}
  H(\kappa \mid v) \ge m - \epsilon_1(n, \epsilon, m)
\end{equation}
in which
\begin{equation}
\epsilon_1 (n, m, \epsilon) = 2(m + \frac{1}{\ln 2}) h(n, \epsilon) + 2 \sqrt{ 2 (m + \frac{1}{\ln2 } ) m h(n, \epsilon) } + m P(\cal{P}
\cap \overline{\cal{L}}_{\epsilon_w}).
\end{equation}
Corollary \ref{cor:smallsphere} now implies that $\epsilon_1$ satisfies the condition \sref{eq:MRPropE2}, which completes the proof.


\end{proof}

\begin{appendix}

\section{Technical Issues}
In this appendix, we present some technical lemma's which were used.
The first lemma concerns the reduction of a trace to a smaller Hilbert space.
\begin{lem}
\label{lm:PartialTrace}
Consider two finite dimensional Hilbert spaces $\cal{H}^A$, $\cal{H}^B$
and the product Hilbert space $\cal{H} = \cal{H}^A \otimes \cal{H}^B$.
Consider two density matrices $\rho^A$ and $\rho^B$ over $\cal{H}^A$ and $\cal{H}^B$
respectively and let $\rho = \rho^A \otimes \rho^B$.
Then for any measurement operator $F$ acting on $\cal{H}$,
there exists a measurement operator $F'_{\rho^B}$ on $\cal{H}^A$
which depends only on $\rho^B$, such that
\begin{equation}
\Tr_{\cal{H}} F \rho = \Tr_{\cal{H}^A} F'_{\rho^B} \rho^A.
\end{equation}
In addition, for any set $\{F_q\}_{q \in Q}$ of measurement operators on $\cal{H}$ such that
\begin{equation}
\sum_{q \in Q}F_q = \mathbf{1}_{\cal{H}},
\end{equation}
we have that $\sum_{q \in Q}F'_{q, \rho^B} = \mathbf{1}_{\cal{H}^A}$
\end{lem}
\begin{proof}
Let $n_A$ and $n_B$ denote the dimension of $\cal{H}^A$ respectively $\cal{H}^B$.
For any four-tuple of integers $(i_A, j_A, i_B, j_B)$ such that $1 \le i_A, j_A \le n_A$ and
$1 \le i_B, j_B \le n_B$, define $e_{i_A, j_A, i_B, j_B} = e^A_{i_A, j_A} \otimes e^B_{i_B, j_B}$,
where $e^A_{i_A, j_A}$ is the $n_A \times n_A$ matrix which has a $1$ at position $(i_A, j_A)$
and zeroes elsewhere and $e^B$ is defined similarly.
Any square matrix $X$ on $\cal{H}$ can be decomposed as
\begin{equation}
X = \sum_{(i_A, j_A, i_B, j_B)} X^{i_A, j_A}_{i_B, j_B} e_{i_A, j_A}^{i_B, j_B}.
\end{equation}
Thus defining
\begin{equation}
X'_{\rho^B} = \sum_{i_A, j_A} e^A_{i_A, j_A} \sum_{i_B, j_B} X^{i_A, j_A}_{i_B, j_B} \Tr_B e^B_{i_B, j_B} \rho^B,
\end{equation}
we see that indeed $\Tr_{\cal{H}}X \rho = \Tr_{A} X'_{\rho^B} \rho^A$.
The fact that $X'_{\rho^B}$ is a nonnegative operator can be seen by taking
$\rho^A = \ket{\alpha}\bra{\alpha}$ for any normalized state $\ket{\alpha}$ in $\cal{H}^A$ and noting that
\begin{equation}
\bra{\alpha}X'_{\rho^B}\ket{\alpha}=\Tr_{A} X'_{\rho^B} \ket{\alpha}\bra{\alpha} = \Tr_{\cal{H}} X \ket{\alpha}\bra{\alpha} \otimes \rho^B  \ge 0.
\end{equation}
The last claim in the lemma can be verified by noting that $(X+Y)'_{\rho^B} = X'_{\rho^B} + Y'_{\rho^B}$ and
$\mathbf{1}'_{\rho^B} = \mathbf{1}_{\cal{H}^A}$, since $\Tr_{B} \rho^B = 1$.
\end{proof}

The following result gives a bound on the success rate of any quantum measurement
which must distinguish between two quantum states. In addition, it shows that performing
collective measurements on random sequences of these two states does not improve the
success rate on individual positions.

\begin{thm}
\label{thm:Distinguish}
Consider two pairs of density matrices $(\rho_a^0, \rho_a^1)$,
for $a =0, 1$.
Denote by $S_a$ the set of eigenvalues $\lambda$ of the matrix $\rho_a^0 - \rho_a^1$ and
define the quantity $\Delta = \max_{a =0, 1}\sum_{\lambda \in S_a} \abs{\lambda}$.
Fix an integer $N$ and a string $\vec{a} \in \Field_2^N$. Let a source emit a sequence of $N$ states, given by a string $g$, where
$g[i] = 0$ when $\rho_{a[i]}^0$ was emitted and $g[i] = 1$ otherwise.
Suppose that at each position both possible states have equal probabilities to occur, i.e. $P(g[i] = 0) = P(g[i] =1) = \frac{1}{2}$.
Consider an arbitrary measurement on the system which gives guesses $\vec{h}$ for $\vec{g}$
for $m \le N$ different positions.
Then the probability that the $m$ guesses are all correct is bounded by
\begin{equation}
P_{\mathrm{success}} \le  \big( \frac{1}{2} + \frac{1}{4} \Delta \big)^m.
\end{equation}

\end{thm}
\begin{proof}
We assume that $\vec{a} = 0$. With the addition of some bookkeeping arguments the proof given below can
be seen to hold for all strings $\vec{a}$.
We model the emission of the source as a state in $H^C \otimes H^S$,
in which $H^C$ is the classical space consisting of bit-strings in $\Field_2^N$
and $H^S = H^{\otimes N}$ is the state space for the emitted quantum states.
Without loss of generality, we shall assume that the $m$ positions for which the guess $h$ is supplied
are the first $m$ positions. Correspondingly, we write $g' \in \Field_2^m$ for the first $m$ bits of $g$.
The measurement determining the guess $h$ and the subsequent announcement of $g'$ can be described by the POVM
\begin{equation}
\{ (h, g') , \Pi^C_{g'} \otimes F_h \},
\end{equation}
since the measurement on the quantum states is independent of the announcement $g'$.
The probability of success thus reads, using Lemma \ref{lm:TraceReduction},
\begin{equation}
P_{\mathrm{success}} = 2^{-m} \sum_{g' \in \Field_2^m} \Tr F_{g'} \rho^S_{g'},
\end{equation}
where
\begin{equation}
\rho^S_{g'} = 2^{m-N} \bigotimes_{k=1}^m \rho_{g'[k]} \otimes \big( \rho_1 + \rho_2 \big)^{\otimes N-m}.
\end{equation}
Splitting $H^S = H^{\otimes m} \otimes H^{rest}$ and using Lemma \ref{lm:PartialTrace} to perform the trace over $H^{rest}$, we obtain
$P_{\mathrm{success}} = \Tr T_m$, in which the trace runs over $H^{\otimes m}$ and $T_m$ is given by
\begin{equation}
T_m = 2^{-m} \sum_{g' \in \Field_2^m} F'_{g'} \rho_{g'}^S.
\end{equation}
Consider the linear space $\cal{W}$ spanned by  words over the alphabet $\{ F^0, F^1, \rho^0, \rho^1 \}$.
For every word $w = w_1 w_2 \ldots w_{2m}$, we define the normalized word $\cal{N}(w)$, which reorders symbols $w_i$ such that
each $F^a$ stands to the left of each $\rho^b$, but that otherwise leaves the ordering invariant. For example,
\begin{equation}
\cal{N} (F^0 \rho^1 F^1 \rho^0) = F^0 F^1 \rho^1 \rho^0.
\end{equation}
For every normalized codeword of the form
\begin{equation}
F^{v[1]} \ldots F^{v[m]} \rho^{w[1]} \ldots \rho^{w[m]}
\end{equation}
we define the corresponding matrix $\cal{M}(w) = F'_v \rho_w$.
These operators can be extended to the complete linear space $\cal{W}$ by simply linearizing.
We recursively define elements in $\cal{W}$ by  $W_0 = \emptyset$ and
\begin{equation}
W_j = (F^0 \rho^0 + F^1 \rho^1 ) W_{j-1} = \big( (F^0 + F^1) \frac{1}{2} ( \rho^0 + \rho^1 ) + (F^0 - F^1) \frac{1}{2} (\rho^0 - \rho^1) \big) W_{j-1}.
\end{equation}
It is not hard to see $T_m = 2^{-m} \cal{M}(\cal{N}(W_m))$.
Write $W_m = (A_0 + A_1)^m$, where $A_0 =(F^0 + F^1) \frac{1}{2} ( \rho^0 + \rho^1 )$ and $A_1 = (F^0 - F^1) \frac{1}{2} ( \rho^0 - \rho^1 )$.
For any $v \in \Field_2^m$, we define the element $A_v = A_{v[0]} A_{v[1]} \ldots A_{v[m]}$. We shall compute
$\Tr \cal{M}(A_v)$.
Without loss of generality, we shall assume that $v = (0, \ldots 0, 1, \ldots 1)$ with $d(v, 0) = s$.
Using Lemma \ref{lm:PartialTrace} to perform the trace over the first $m-s$ positions, we are left with
\begin{equation}
\Tr \cal{M}(A_v) = \sum_{w \in \Field_2^s} \epsilon(w) 2^{-s} \Tr F''_{w} (\rho^0 - \rho^1)^{\otimes s},
\end{equation}
where $F''_{w} = \sum_{ r \in \Field_2^{m-s}} (F'_{rw})'_{(\frac{1}{2}(\rho^0 + \rho^1))^{\otimes m-s}}$
and where $\epsilon(w) = (-1)^{d(w, 0)}$ is a $\pm 1 $ valued function.
Pass to a basis for which $\rho^0 - \rho^1$ is diagonal and let $F'''_{w}$ be $F''_w$ in this basis.
Note that we have $\sum_{w \in \Field_2^s} F'''_{w} = \mathbf{1}_{H^s}$ and that each $F'''_w$ is a measurement operator,
which means that all the diagonal elements $d_{ii}$
of $\sum_{w \in \Field_2^s} \epsilon(w) F'''_{w}$ have norm $\abs{d_{ii}} \le 1$.
In particular, this means
\begin{equation}
\abs{\Tr \cal{M}(A_v)} \le 2^{-s} \Delta^s.
\end{equation}
We can thus compute, summing $\Tr \cal{M}(A_v)$ over all $v$,
\begin{equation}
\Tr T_m \le 2^{-m} (1 + \frac{\Delta}{2})^m,
\end{equation}
which proves the claim.

\end{proof}

\noindent The following two results are standard bounds on the tails of binomial distributions.

\begin{lem}
  \label{lm:binomialTail}
  Let $p$, $r$ and $t$ be positive numbers such that $0 < r \le p < p+t < 1$. Let $n_r$ and $n_p$ be two positive
  integers and define the set $V = \left\{ (i_r, i_p) \in \Numbers_0 \times \Numbers_0 \mid i_r + i_p \ge (p +t ) n \right\}$,
  where $n = n_r  + n_p$. Then
  \begin{equation}
    \sum_{(i_r, i_p) \in V}{ { n_p \choose i_p} { n_r \choose i_r } p^{i_p} (1-p)^{n_p - i_p} r^{i_r}(1-r)^{n_r - i_r}   } \le e^{-2t^2n}.
  \end{equation}
\end{lem}
\begin{proof}
  For simplicity, we define  $q = 1-p$, $s = 1-r$, $k = \lceil (p+t)n \rceil$ and write
  \begin{equation}
    S = \sum_{(i_r, i_p) \in V}{ { n_p \choose i_p} { n_r \choose i_r } p^{i_p} (1-p)^{n_p - i_p} r^{i_r}(1-r)^{n_r - i_r}   }.
  \end{equation}
  Then for any $x \ge 1$, one has
  \begin{equation}
    \begin{array}{lcl}
      S &  \le & \sum_{(i_r, i_p) \in V}{ { n_p \choose i_p} { n_r \choose i_r } p^{i_p} (1-p)^{n_p - i_p}x^{i_p - k} r^{i_r}(1-r)^{n_r - i_r}x^{i_r - k}   } \\
      & \le &  \sum_{0 \le i_p \le n_p}{\sum_{ 0 \le i_r \le n_r}{ { n_p \choose i_p} { n_r \choose i_r } p^{i_p} (1-p)^{n_p - i_p} r^{i_r}(1-r)^{n_r - i_r}   } } \\
      & = & \frac{1}{x^k} (q+px)^{n_p} (s + rx)^{n_r} \le \frac{1}{x^{(p+t)n}} (q + px)^{n_p}(s+rx)^{n_r} \le \frac{1}{x^{(p+t)n}} (q + px)^{n}, \\
    \end{array}
  \end{equation}
  where we have used $s + rx \le q + px$ in the last inequality.
  Fixing $x = \frac{q(p+t)}{p(q-t)} \ge 1$, we obtain
  \begin{equation}
    S \le \Big[ \big( \frac{p}{p+t} \big) ^{p+t} \big( \frac{q}{q-t}\big) ^{q-t} \Big] ^n   .
  \end{equation}
  Define the function
  \begin{equation}
    g(t) = \ln \Big[ \big( \frac{p}{p+t} \big) ^{p+t} \big( \frac{q}{q-t}\big) ^{q-t} \Big].
  \end{equation}
  It is easy to see that $g$ is $C^{\infty}$ on $[0, q]$, so we may employ Taylor's formula to get
  \begin{equation}
    g(t) = g(0) + t g'(0) + \int_{0}^{t}{g''(u) (t-u) du}.
  \end{equation}
  Notice that $g(0) = g'(0) = 0$ and $g''(u) = - \frac{1}{(p+u)(q-u)} \le -4$ for any $u \in [0, q]$.
  Therefore $g(t) \le -2 t^2$ from which the statement follows.
\end{proof}

\begin{cor}
  \label{cr:binomialTail}
  Let $p$, $r$ and $t$ be positive numbers such that $0 <  r-t \le r \le p < 1$. Let $n_r$ and $n_p$ be two positive
  integers and define the set $V = \left\{ (i_r, i_p) \in \Numbers \times \Numbers \mid i_r + i_p \le (r -t ) n \right\}$,
  where $n = n_r  + n_p$. Then
  \begin{equation}
    \sum_{(i_r, i_p) \in V}{ { n_p \choose i_p} { n_r \choose i_r } p^{i_p} (1-p)^{n_p - i_p} r^{i_r}(1-r)^{n_r - i_r}   } \le e^{-2t^2n}.
  \end{equation}
\end{cor}
\begin{proof}
This follows immediately from Lemma \ref{lm:binomialTail} by  making the substitutions $r \to 1-r$,
$p \to 1-p$ and recalling that ${ n \choose k} = { n \choose n-k}$.
\end{proof}

\noindent This next result is a classic result which follows directly from the shape of the logarithm.

\begin{lem}[Jensen]
\label{lm:Jensen}
Consider real numbers $a_1, \ldots a_m$ and $b_1, \ldots b_m$ and suppose that $0 \le a_1 \le 1$, $b_i > 0$ and
$\sum_{i =1}^m a_i = 1$. Then
\begin{equation}
\sum_{i=1}^m a_i \log_2 b_i \le \log_2 \sum_{i=1}^m a_i b_i.
\end{equation}
\end{lem}

\begin{lem}
\label{lm:SmallProbability}
Let $\mu > 0$ be a strictly positive real number. Let $y$ be a random variable taking values in a set $\cal{Y}$ and
let $\{ a_y \}_{y \in \cal{Y}}$ be a set of $\abs{\cal{Y}}$ real nonnegative numbers such that $\sum_{ y \in \cal{Y}} a_y \le \mu$.
Let $q$ be a strictly positive number and define the subset $\cal{S} \subseteq \cal{Y}$ by
\begin{equation}
\cal{X} = \{ y \in \cal{Y} \mid a_y \le \mu q P_y(y).
\end{equation}
Then $P_y(\cal{X}) \ge 1 - \frac{1}{q}$.
\end{lem}
\begin{proof}
Assume to the contrary that $P_y(\cal{Y} \setminus \cal{X}) > \frac{1}{q}$.
Then
\begin{equation}
\sum_{y \in \cal{Y}} a_y \ge \sum_{y \in \cal{Y} \setminus \cal{X}}{a_y} > \mu q \sum_{y \in \cal{Y} \setminus \cal{X}}{P_y(y)} = \mu q P(\cal{Y} \setminus \cal{X}) > \mu,
\end{equation}
which is a contradiction.
\end{proof}


\section{Error Correcting Codes}
\label{sc:aperr}
Consider two integers which satisfy $1 \le k \le n$ and let $G$ be a $k \times n$ binary matrix with linearly
independent rows. Define the set $\cal{S}(G) = \{ w \in \Field_2^n \mid w = v G \hbox{ for some } v \in \Field_2^k\}$,
which is a linear subspace of $\Field_2^n$ of dimension $k$. Letting $d_G =\min_{g \in \cal{S}(G)}d(g,0)$
be the the minimum weight of strings in $\cal{S}(G)$, we say that the set $\cal{S}(G)$ is a $(n, k)$ linear code
with minimum distance $d_G$.
For any such matrix $G$ the map $Enc_{G}: \Field_2^k \to \Field_2^n$ which sends $v \to vG$ is an inclusion from $\Field_2^k$ into $\Field_2^n$
and can be used to encode messages in $\Field_2^k$ into strings in the larger space $\Field_2^n$.
The intuitive idea of an error correcting code is to use the redundancy in this encoding to protect any encoded string
from bitflips in a small number of positions. This is usually done by means of minimal distance decoding, that is,
for any string $s \in \Field_2^n$, one defines $Dec_{G}(s) \in \Field_2^k$ to be a string $s_{org}$ that minimizes $d(s_{org}G, s)$.
Let $t$ be any positive integer satisfying $2t+1 \le d_G$ and let $e \in \Field_2^n$ be an arbitrary string with weight $d(e, 0) = t$.
Since $d(Enc(s_{org}) , Enc(s'_{org})) = d((s_{org} \ominus s'_{org})G,0) \ge d_G \ge 2t+1$ whenever $s_{org} \neq s'_{org}$, we see that
we must have $Dec_{G}( Enc(s_{org}) \oplus e) = s_{org}$ for any message $s_{org} \in \Field_2^k$. We thus see that the decoding
scheme functions correctly whenever the number of bitflips which have occurred on the encoded string does not exceed
$t_{max} = \lfloor \frac{d_G-1}{2} \rfloor$
and we correspondingly say that the code $\cal{S}(G)$ is an error correcting code which can correct $t_{\max}$ errors.

It can be shown that there exists a binary $(n-k) \times n$ matrix $H$ for which $Hg =0$ if and only if $g \in \cal{S}(G)$.
This matrix is called the parity check matrix of the code $\cal{S}(G)$. For a given $x \in \Field_2^n$, we call $s = Hx$ the
syndrome of $x$. Notice that whenever two strings $x, x'$ share the same syndrome $s$, we have $H(x \ominus x') = Hx \ominus Hx' = s \ominus s = 0$
and hence $x \ominus x' \in \cal{S}(G)$. We can exploit this fact by defining a decode function $Dec^s_G: \Field_2^n \to \Field_2^n$
which computes $Dec^s_G(y) = x$ for any $x$ which satisfies $Hx = s$ and $d(y, x) = \min \{ d(y, x') \mid Hx' = s\}$.
Using the same arguments as above, it can be seen that for any error string $e$ with $d(e, 0) \le t_{\max}$, we have
$Dec^{Hx}_G(x \oplus e) = x$. This fact was used to prove that Alice and Bob share the same secret key at the end of the protocol
if $d_{S_{\cal{K}}}(g, h) \le \delta_{\cal{P}} + \epsilon$.

The next basic results give some minimal bounds on the efficiency of error correcting codes and were used to establish
the worst-case asymptotic rate of key generation \sref{eq:asympRate}.

\begin{lem}[Gilbert-Varshamov]
For any strictly positive integers $n, r,t$ which satisfy
\begin{equation}
\label{eq:gvCond}
2^{r + 1} > \sum_{i=0}^{2t} { n \choose k},
\end{equation}
there exists a linear $(n, n-r)$ code which can correct $t$ errors.
\end{lem}
\begin{proof}
We will construct a $n-r \times n$ generator matrix $G$ of a code which has minimum distance $d$ which satisfies $d \ge 2t+1$
and can hence correct $t$ errors.
Set $v_1$ to be an arbitrary vector from $\Field_2^n$ which has weight $2t+1$
and iteratively choose vectors $v_i$ such that for every $i$ the set $\{v_1, \ldots, v_i \}$ is linearly independent
and all the nonzero vectors in $\mathrm{span}( \{v_1, \ldots, v_i \})$ have a weight of at least $2t+1$.
For any $i$, this is possible if there are still vectors in $\Field_2^n$ outside the spheres
of radius $2t$ around the $2^{i-1}$ codewords in $\mathrm{span}( \{v_1, \ldots v_i \})$.
Since each sphere of radius $2t$ contains $\sum_{i=0}^{2t} {n \choose k}$ points,
\sref{eq:gvCond} implies that we can construct $v_1, \ldots, v_{n-r}$ in this way.
The claim immediately follows if we let $v_1, \ldots,  v_{n-r}$ be the rows of $G$.
\end{proof}

\begin{lem}[{\cite[Corollary 9]{MacWSlo}}]
\label{lm:binsum}
For any $0 < \mu < \frac{1}{2}$ and for any integer $n$, we have
\begin{equation}
\sum_{k=0}^{\lfloor \mu n \rfloor} { n \choose k} \le 2^{nH_2(\mu)},
\end{equation}
where $H_2(\mu)$ is the binary entropy function $H_2(\mu) = - ( \mu \ln \mu + (1-\mu) \ln (1- \mu) )$.
\end{lem}

\noindent Combining the previous two lemma's gives us the following asymptotic expression of the Gilbert-Varshamov bound.
\begin{cor}
\label{cr:asympGV}
Fix  $0 < \delta < \frac{1}{4}$. Then for every $n$ there exists
an $(n, n-r)$ error correcting code that can correct $\lfloor \delta n \rfloor$ errors for some $r$ which satisfies
\begin{equation}
\frac{r}{n} \le H_2( 2\delta).
\end{equation}
\end{cor}

\begin{lem}
\label{lm:privAmp}
Fix three positive integers $r$, $n$ and $d_{\min}$ and
consider an arbitrary $r \times n$ binary matrix $F$ with linearly independent rows.
Let $\cal{F}$ be the set containing the $r$ rows of the matrix $F$.
Suppose that $2^{n-r-m + 1} > \sum_{i=0}^{d_{min} - 1} { n \choose i}$.
Then there exists a set $W$ containing $m$ vectors in $\Field_2^n$ such that the set $W \cup \cal{F}$ is
linearly independent and for every $v$ in
the set $\mathrm{span}(W \cup \cal{F}) \setminus \mathrm{span}(\cal{F})$, the inequality
$d(v) \ge d_{\min}$ holds for the weight $d(v) = d(v, 0)$.
\end{lem}
\begin{proof}
Let $S$ be the set of vectors $v$ in $\Field_2^N$ which have weight $d(v) < d_{min}$.
Then $\abs{S} = \sum_{i=0}^{d_{min} -1} {n \choose i}$.
We inductively define a sequence of sets $W_i$ for $0 \le i \le m$ with the property
that $W_i$ contains $i$ distinct vectors from $\Field_2^n$, the set $W_i \cup \cal{F}$ is linearly independent
and $d(v) \ge d_{min}$ for every $v \in \mathrm{span}(W_i \cup \cal{F}) \setminus \mathrm{span}(\cal{F})$.
Let $W_0 = \emptyset$ which can easily be seen to satisfy the above properties.
For any $0 \le i < m$, let $\cal{W}_i = \mathrm{span}(W_i \cup \cal{F})$.
Since there are $2^{n-r -i} \ge 2^{n-r-m + 1} > \abs{S}$ distinct cosets of $\cal{W}_i$ in $\Field_2^n$, there is
at least one such coset which has empty intersection with $S$. Let $w_{i+1}$ be a representative of such a coset and
define $W_{i + 1} = \{w_{i+1} \} \cup W_i$. It is easy to see that if $W_i$ satisfies the properties mentioned
above, then this also holds for $W_{i+1}$ and the set $W_m$ can thus indeed be defined. This completes the proof.
\end{proof}

\end{appendix}



\end{document}